\documentclass[fleqn,usenatbib]{mnras}

\usepackage{newtxtext,newtxmath}
\usepackage[T1]{fontenc}
\usepackage{ae,aecompl}

\usepackage{graphicx}	% Including figure files
\usepackage{multicol}        % Multi-column entries in tables
\usepackage{bm}		% Bold maths symbols, including upright Greek
\usepackage{pdflscape}	% Landscape pages
\usepackage{url}
\title[Gaia EDR3 Bar Resonances]{Galactic Bar Resonances Inferred from Kinematically Hot Stars in {\it Gaia} EDR3}

\author[D. Kawata et al.]{
Daisuke Kawata$^{1}$\thanks{E-mail: d.kawata@ucl.ca.uk}, 
Junichi Baba$^{2,3}$, 
Jason A. S. Hunt$^{4}$,
Ralph Sch\"onrich$^{1}$,
\newauthor{
Ioana Ciuc\u{a}$^{1,5,6}$, 
Jennifer Friske$^{1,7}$,
George Seabroke$^{1}$
Mark Cropper$^{1}$,
}
\\
$^{1}$Mullard Space Science Laboratory, University College London, Holmbury St. Mary, Dorking, Surrey, RH5 6NT, UK\\
$^{2}$National Astronomical Observatory of Japan, Mitaka, Tokyo 181-8588, Japan\\
$^{3}$SOKENDAI (The Graduate University for Advanced Students), Shonan Village, Hayama, Kanagawa 240-0193, Japan.\\
$^{4}$Center for Computational Astrophysics, Flatiron Institute, 162 5th Av., New York City, NY 10010, USA\\
$^{5}$Research School of Astronomy and Astrophysics, Mount Stromlo Observatory, Cotter Road, Weston Creek, ACT 2611, Canberra, Australia\\
$^{6}$Centre of Excellence for Astrophysics in Three Dimensions (ASTRO-3D), Australia\\
$^{7}$Ludwig-Maximilians-Universit{\"a}t, Fakult{\"a}t f{\"u}r Physik Schellingstr. 4, 80799 M{\"u}nchen, Germany\\
}
\date{Accepted XXX. Received YYY; in original form ZZZ}

\pubyear{2018}

\begin{document}
\label{firstpage}
\pagerange{\pageref{firstpage}--\pageref{lastpage}}
\maketitle
% * <babajn2000@gmail.com> 2018-03-05T07:51:15.330Z:
\begin{abstract}
Using a numerical simulation of an isolated barred disc galaxy, we first demonstrate that the resonances of the inner bar structure induce more prominent features in the action space distribution for the kinematically hotter stars, which are less sensitive to the local perturbation, such as the transient spiral arms. Then, we analyse the action distribution for the kinematically hotter stars selected from the {\it Gaia}~EDR3 data as the stars with higher values of radial and vertical actions. We find several resonance features, including two new features, in the angular momentum distribution similar to what are seen in our numerical simulations. We show that the bar pattern speeds of about $\Omega_{\rm bar}\sim34$~km~s$^{-1}$~kpc$^{-1}$ and 42~km~s$^{-1}$~kpc$^{-1}$ explain all these features equally well. The resonance features we find correspond to the inner 4:1, co-rotation, outer 4:1, outer Lindblad and outer 4:3 (co-rotation, outer 4:1, outer Lindblad, outer 4:3 and outer 1:1) resonances, when $\Omega_{\rm bar}\sim34$ (42) km~s$^{-1}$~kpc$^{-1}$ is assumed. 
\end{abstract}


%%%%%%%%%%%%%%%%%%%%%%%%%%%%%%%%%%%%%%%%%%%%%%%%%%

%%%%%%%%%%%%%%%%% BODY OF PAPER %%%%%%%%%%%%%%%%%%

\section{Introduction}
\label{sec:intro}

The central few kpc of the Milky Way show a prominent bar structure \citep[e.g.][]{bhg16}. The solidly rotating bar components affect the radial and rotational velocity distribution of the Galactic disc stars, and the presence of groups of stars moving with particular radial and rotational velocities in the Solar neighbourhood can be attributed to the bar \citep{wd00}. \cite{Dehnen99} suggested that the Hercules stream, which is a group of stars rotating slower and moving outward in the disc, is caused by the outer Lindblad resonance (OLR hereafter) of the bar being just inside of the Sun's orbital radius. If true, this allows us to derive the pattern speed of the bar \citep{Dehnen99,Monari+17,Fragkoudi+19}, and it is found to be fast \citep[e.g. 53 km s$^{-1}$ kpc$^{-1}$;][]{Dehnen99}.

{\it Gaia} data release 2 \citep[DR2,][]{Gaia+Brown+18} revolutionised our view of the kinematic structure of stars not only in the Solar neighbourhood but also for several kpc across the Galactic disc \citep{Gaia+Katz18Disc,Antoja+18,Kawata+18,Friske+Schoenrich19}. It is well complimented by near-infrared photometric surveys, such as {\it VISTA} Variables in the Via Lactea \citep[VVV;][]{Minniti+Lucas+Emerson+10}, and ground-based spectroscopic surveys, such as Bulge Radial Velocity Assay \citep[BRAVA;][]{Kunder+Koch+Rich+12}, the Abundances and Radial velocity Galactic Origins Survey \citep[ARGOS;][]{Freeman+Ness+13,Ness+Freeman+Athanassoula+13} and the Apache Point Observatory Galactic Evolution Experiment \citep[APOGEE;][]{Majewski+Schiavon+Frinchaboy+17}, which revealed the detailed stellar structure and line-of-sight velocities within the Galactic bar itself. These observations were directly compared with theoretical models \citep[e.g.][]{Shen+Rich+Kormendy+10,Portail+15}, which suggest a slower pattern speed of the bar than is found when assuming the OLR to be just inside of the Solar radius. Recently, both the gas dynamics \citep{Sormani+15c} and stellar dynamics from {\it Gaia}~DR2, combined with VVV and APOGEE data \citep{Sanders+Smith+Evans19,Bovy+Leung+Hunt+19} are converging on a value for the bar pattern speed of around $40$~km~s$^{-1}$~kpc$^{-1}$.

Interestingly, the pattern speed of $40$~km~s$^{-1}$~kpc$^{-1}$ can explain the Hercules stream with the outer 4:1 resonance \citep{Hunt+Bovy18}. However, for such a pattern speed there should be a clear OLR feature in the kinematics just outside of the Solar neighbourhood \citep{Hunt+Bub+Bovy+19,Trick+Fragkoudi+Hunt+21}. \citet{Perez-Villegas+17,Monari+Famaey+Siebert+Wegg+Gerhard19} suggested that the co-rotation (CR hereafter) is attributed to the Hercules stream, which explains the other features better \citep[see a comprehensive discussion in][]{Trick+Fragkoudi+Hunt+21}. Most models that attempt to create the Hercules stream from the CR alone show that the effect on the local velocity distribution is significantly weaker than the observed data \citep[e.g.][]{Binney18a,Hunt+Bovy18}. This is reconcilable with the addition of spiral structure \citep{Hunt+Hong+Bovy+18, Hunt+Bub+Bovy+19}. Alternatively, while previous studies assume a bar which rotates with a fixed pattern speed, \citet{Chiba+Friske+Schoenrich20,Chiba+Schoenrich21} demonstrated that the observed detailed features of the stellar phase space distribution are better explained by the scenario that the Galactic bar is slowing down, and their CR reproduces the kinematics of the Hercules stream without requiring spiral structure. Hence, the pattern speed of the bar is still in-debate and requires more data and theoretical modelling studies.

Confronting the data with various theoretical models has made us realise that the kinematic features observed in the Solar neighbourhood can be explained by several different pattern speeds of the Galactic bar \citep[e.g.][]{Trick+Coronado+Rix+19,Trick+Fragkoudi+Hunt+21,Hunt+Bub+Bovy+19}, because the bar can induce similar features with different resonances, such as the co-rotation resonance \citep[e.g.][]{Perez-Villegas+17,D'Onghia+Aguerri20,Chiba+Friske+Schoenrich20}, the outer 4:1 resonance \citep[simply 4:1R hereafter;][]{Hunt+Bovy18}, the OLR, or other higher order resonances \citep[e.g.][]{Monari+Famaey+Siebert+Wegg+Gerhard19,Asano+20}. While kinematic structure induced by different order resonances will vary significantly over Galactic scales, it is non-trivial to identify causation with Solar neighborhood data, and strong features like the Hercules stream can be explained in multiple ways. In addition, it is further complicated by the fact that similar phase space features can also be caused by transient spiral arms \citep{DeSimone+Wu+Tremaine04,qdbmc11, Hunt+Hong+Bovy+18,Fujii+2019}, and the influence of dwarf galaxies such as Sagittarius \citep[e.g.][]{LMJG19,Khanna+19}, and in many cases these will overlap in kinematic space.

In particular, transient spiral arms can have a systematic effect on the velocity field around the spiral arms \citep[e.g.][]{sb02,gkc12a,bsw13,khgpc14} and radial migration causes nontrivial effects on the orbital phase \citep[e.g.][]{gkc15}. Hence, the velocity field and phase angles of orbits derived from an asymmetric potential are heavily affected by the transient spiral arms. In addition, the phase angles currently suffer from significant selection effects \citep{Hunt+Johnston+Pettitt+20,Trick21}. Hence, in this paper we focus on kinematically `hotter' stars in the Galactic disc, which can be defined as stars with larger actions. We consider that kinematically hotter stars are less affected by the transient spiral arms.
% which are much weaker features than the Galactic bar {\it citation needed}. 
% DK: perhaps it is difficult to define the strength. Anyway, this is a working assumption to be proven. So, it is ok to delete this sentence. 
On the other hand, the Galactic bar resonances are relatively strong and long lived effects, and can affect the kinematically hot stars as well \citep{Binney18a}. 

In Section~\ref{sec:sim-reso} we first demonstrate that this working assumption is valid, based on the results of an $N$-body/SPH simulation. Then, Section~\ref{sec:gedr3-reso} shows the action distribution of stars in the recently released {\it Gaia}~early data release 3 \citep[{\it Gaia}~EDR3;][]{Gaia+Brown+21}. Using the results from $N$-body/SPH simulation as a prior, we discuss which observed action space features of stars correspond to which resonances of the Galactic bar. Summary and discussion of this study are presented in Section~\ref{sec:sum}.

\section{Bar Resonances Features in {\it N}-body/SPH Simulation}
\label{sec:sim-reso}

We utilise the $N$-body/SPH simulation of a Milky Way-like galaxy presented in \citet{Baba+Kawata20} and \citet{Baba+Kawata+Schoenrich21} to study the features expected to arise from the Galactic bar resonances. This simulation is an isolated disc galaxy, consisting of gas and stellar discs and a classical bulge, evolved self-consistently within a rigid dark matter halo. It includes gas radiative cooling, far-ultraviolet heating, star formation and stellar feedback \citep{Saitoh+2008,Baba+2017}. The gas and stellar particle masses are about $9.1\times10^3$~M$_{\sun}$ and $3\times10^3$~M$_{\sun}$, respectively, and the softening length is set to be 10~pc. 

We use a snapshot at $t=7$~Gyr of this simulation, where there is a clear bar whose pattern speed is around $\Omega_{\rm bar}\sim40$~km~s$^{-1}$. The bar also has an X-shaped boxy inner bar/bulge, and there are several transient spiral arms \citep{Baba+Kawata20}. The bar pattern speed, $\Omega_{\rm bar}$, is measured by the change of the phase angle of $m=2$ Fourier mode, and the time evolution of $\Omega_{\rm bar}$ is shown in Fig.~3 of \citet{Baba+Kawata+Schoenrich21}. Interestingly, $\Omega_{\rm bar}$ fluctuates in the time scale of about 100~Myr, likely due to the interaction with the spiral arm features \citep{Wu+Pfenniger+Taam16,Hilmi+Minchev+Buck+20}. We have selected the star particles with galactocentric radius $3<R<18$~kpc to avoid analysing too many particles in the inner region and with height $|z|<0.5$~kpc from the mid-plane of the disc. The relatively broad vertical region is selected to include kinematically hot disc particles. We select the wide radial range to cover many different resonances. Since we focus on the actions and orbital frequencies only, we use all the particles irrespective of their azimuthal angle position. 

The number of particles are peaked around 8~kpc, because the density profile of the disc falls exponentially in the outer disc, and the area of the disc becomes smaller at smaller radii. To compensate for the decrease in the number of particles available for our analysis in the inner and outer disc, we weight the contribution of the particles from the inner and outer disc to make the weighted number of our particle sample at different radii constant.
To compute the weight for each star, we first count the number of particles, $N_{\rm p}(R_{\rm i})$, in the 64 radial bins within our sample radial range of $3<R<18$~kpc, and compute the weight at the centre of each bin by $w_{\rm i}=\max(N_{\rm p})/N_{\rm p}(R_{\rm i})$, where $\max(N_{\rm p})$ is the number of particles in the bin containing the maximum number of particles. Then, we compute the weight for each particle, depending on their radius by a linear interpolation of the weights at the centre of the radial bins.
Although this has a negligible effect on our $N$-body/SPH simulation results, we find that this is important for the observational data we analyse in the next section. 

We compute the actions and orbital frequencies of the selected star particles using {\tt AGAMA} \citep{Vasiliev_AGAMA19} under the approximated gravitational potential of the $N$-body/SPH simulation snapshot evaluated by {\tt AGAMA} itself. In this paper, we focus on the radial action, $J_{\rm R}$, vertical action, $J_{\rm z}$, and azimuthal action, which is angular momentum, $L_{\rm z}$, and the radial frequency, $\Omega_{\rm R}$, vertical frequency, $\Omega_{\rm z}$, and azimuthal frequency, $\Omega_{\rm \phi}$.

\begin{figure}
 \includegraphics[width=\hsize]{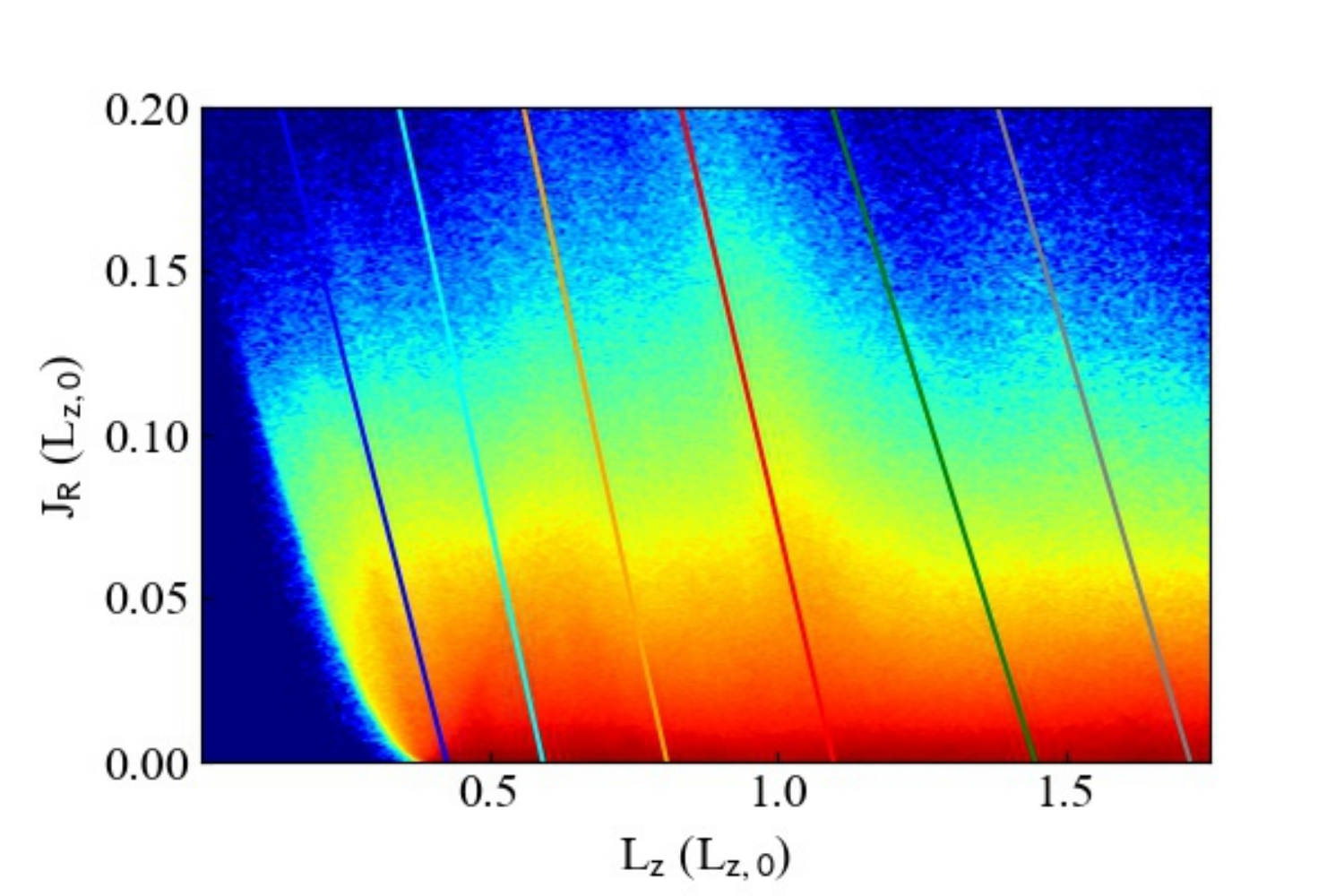}
\caption{The distribution of the angular momentum, $L_{\rm z}$, and radial action, $J_{\rm R}$, for the selected star particles of our $N$-body/SPH simulation. The blue, cyan, orange, red, green and grey solid lines indicate the i4:1R, CR, 4:1R, OLR, 4:3R and 1:1R, respectively.}
\label{fig:lzjr-sim}
\end{figure}

Fig.~\ref{fig:lzjr-sim} shows the distribution of $L_{\rm z}$-$J_{\rm R}$ for the selected particles in our simulations. The actions are normalised by the circular velocity of $V_{\rm circ}=197$~km~s$^{-1}$ at 8~kpc of the simulation. As shown with the {\it Gaia}~DR2 data \citep[e.g.][]{Trick+Coronado+Rix+19,Trick+Fragkoudi+Hunt+21,Hunt+Bub+Bovy+19}, our simulation also shows several strong ridge features. These features are considered to be caused by the resonances of the bar and transient spiral arms \citep[e.g.][]{Hunt+Bub+Bovy+19,Trick+Fragkoudi+Hunt+21}, as discussed in Section~\ref{sec:intro}. The resonances of the bar with pattern speed $\Omega_{\rm bar}$ are defined by the condition of
\begin{equation}
  \Omega_{\rm bar} = \Omega_{\rm \phi} + \frac{l}{m} \Omega_{\rm R},
  \label{eq:resonances}
\end{equation}
where $l$ and $m$ are integer values \citep[e.g.][]{Binney+Tremaine08}. The condition of $\Omega_{\rm bar}=\Omega_{\rm \phi}$, i.e. the rotation frequency of the stars is equal to the bar pattern frequency, is known as the CR. The resonances with $(m, l)=(4, -1), (4, 1)$, $(2, 1)$, $(4, 3)$ and $(1, 1)$ are called the inner 4:1 resonance (i4:1R hereafter), 4:1R, OLR, outer 4:3 resonance (4:3R hereafter) and outer 1:1 resonance (1:1R hereafter), respectively. We use $\Omega_{\rm \phi}$ and $\Omega_{\rm R}$ from {\tt AGAMA} to select the particles around i4:1R, CR, 4:1R, OLR, 4:3R and 1:1R. Then, we apply the robust linear regression, {\tt RANSAC Regressor} in {\tt scikit-learn} \citep{scikit-learn}, to identify the i4:1R, CR, 4:1R, OLR and 1:1R, which are indicated by the blue, cyan, orange, red, green and grey lines in Fig.~\ref{fig:lzjr-sim}, respectively. 

Fig.~\ref{fig:lzjr-sim} shows that the lines of 4:1R and OLR are well aligned with the two major ridges, which are more clear in the higher $J_{\rm R}$, e.g.\ $J_{\rm R}>0.05L_{\rm z,0}$. The ridges are also seen in the lower $J_{\rm R}$. However, for these kinematically colder stars, there are many other features, and these two ridges are not as dominant as what we can see at higher $J_{\rm R}$. 

\begin{figure}
 \includegraphics[width=\hsize]{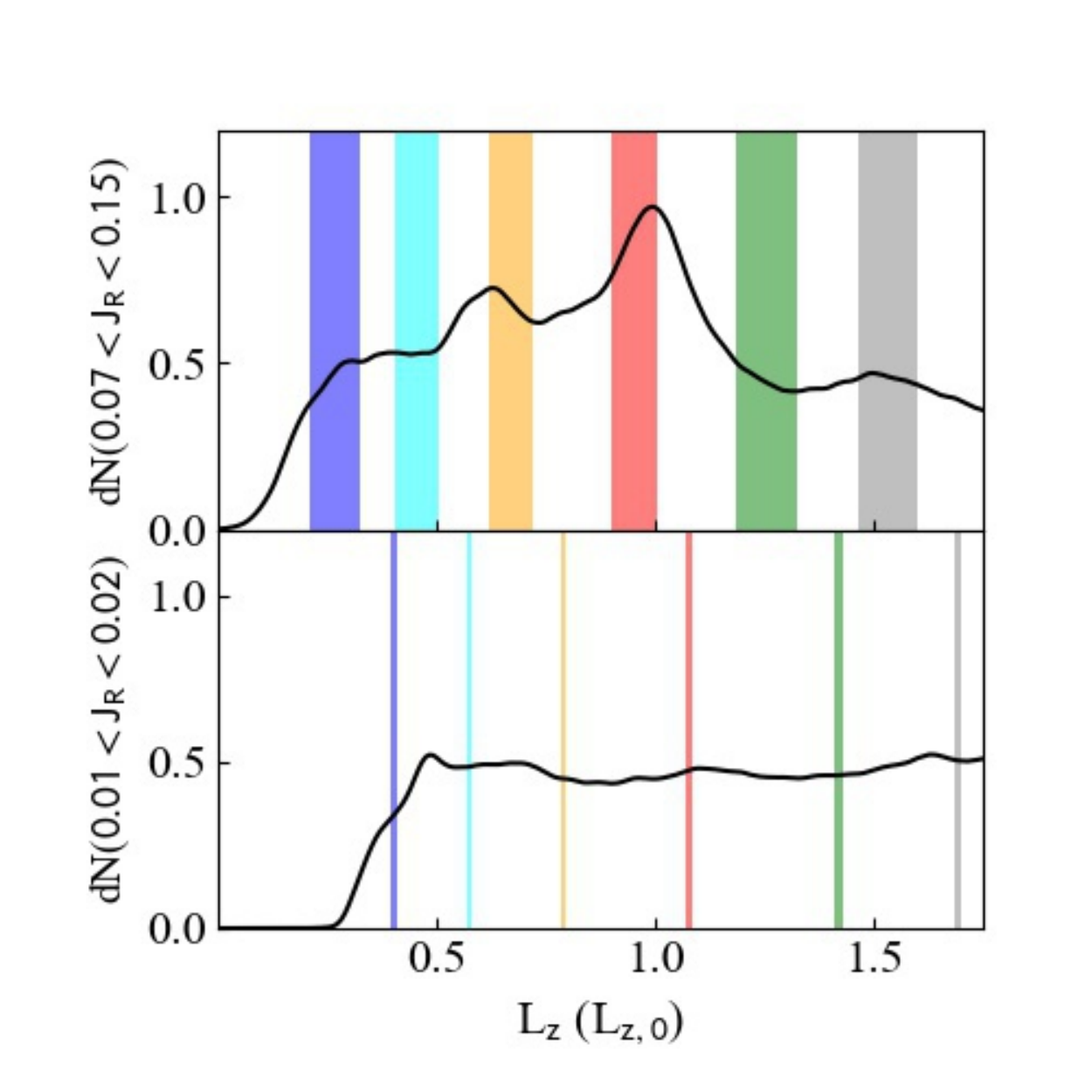}
\caption{The KDE distribution of $L_{\rm z}$ for the star particles with different ranges of $J_{\rm R}$, $0.07<J_{\rm R}<0.15$ (upper) and $0.01<J_{\rm R}<0.02$ (lower). The vertical bands highlighted with blue, cyan, orange, red, green and grey respectively indicate the range of i4:1R, CR, 4:1R, OLR, 4:3R and 1:1R in the $J_{\rm R}$ range of each panel measured from Fig.~\ref{fig:lzjr-sim}.}
\label{fig:lzhist-sim}
\end{figure}

Fig.~\ref{fig:lzhist-sim} demonstrates that the two ridges associated to the 4:1R and OLR are more prominent for the stars with the higher $J_{\rm R}$. As suggested from the cosmological simulations of the barred galaxies in \citet{Fragkoudi+Grand+Pakmor+20}, the OLR ridge is most prominent. The upper panel of Fig.~\ref{fig:lzhist-sim} shows the distribution of $L_{\rm z}$ for the stars with $0.07<J_{\rm R}(L_{\rm z,0})<0.15$. The distributions of $L_{\rm z}$ are computed with {\tt scikit-learn} Kernel Density Estimation \citep[KDE;][]{scikit-learn} with a Epanechnikov kernel with the kernel size of 0.03. The two peaks are prominent and co-located with the 4:1R and OLR highlighted with the orange and red bands, respectively. On the other hand, the lower panel shows the $L_{\rm z}$ distribution for kinematically colder, $0.01<J_{\rm R}(L_{\rm z,0})<0.02$, stars. The distribution is much flatter, because many features are overlapping with each other. Hence, we consider that the strong resonance features, such as the 4:1R and OLR, are easier to identify in the kinematically hot stars. 

The mechanism causing the ridge features around the resonances is still in-debate \citep[e.g.][]{Trick+Fragkoudi+Hunt+21}. The resonance scattering \citep[e.g.][]{lbk72,jas10} could be  responsible for the strong features in the higher $J_{\rm R}$. The orbital trapping \citep[e.g.][]{Monari+Famaey+Fouvry+Binney17,Binney18a,Chiba+Friske+Schoenrich20,Chiba+Schoenrich21} could also be responsible for the feature along the resonances, and the features in higher $J_{\rm R}$ could be due to a higher number of stars being trapped in the resonance, and a higher $J_{\rm R}$ tail being more clear. The velocity fields of stars are also expected to be affected by these mechanisms. However, as discussed in Section~\ref{sec:intro}, we consider that the transient spiral arms can wash out these velocity features. In fact, although it is not shown here, we find that the velocity fields of our $N$-body/SPH simulation are not similar to what are shown in test particle simulations without transient spiral arms. The velocity fields around the spiral arms show the systematic motions due to the transient spiral arms \citep[e.g.][]{khgpc14,Grand+Springel+Kawata+16} to be the dominant effects on the velocity field rather than the bar resonances. Hence, we do not look at the velocity fields or the orbital phase angles, but focus on the action distribution only. Selecting particles with higher action also helps reduce the effect of the transient arms \citep[e.g.][]{sss12}. 

\begin{figure*}
 \includegraphics[width=\hsize]{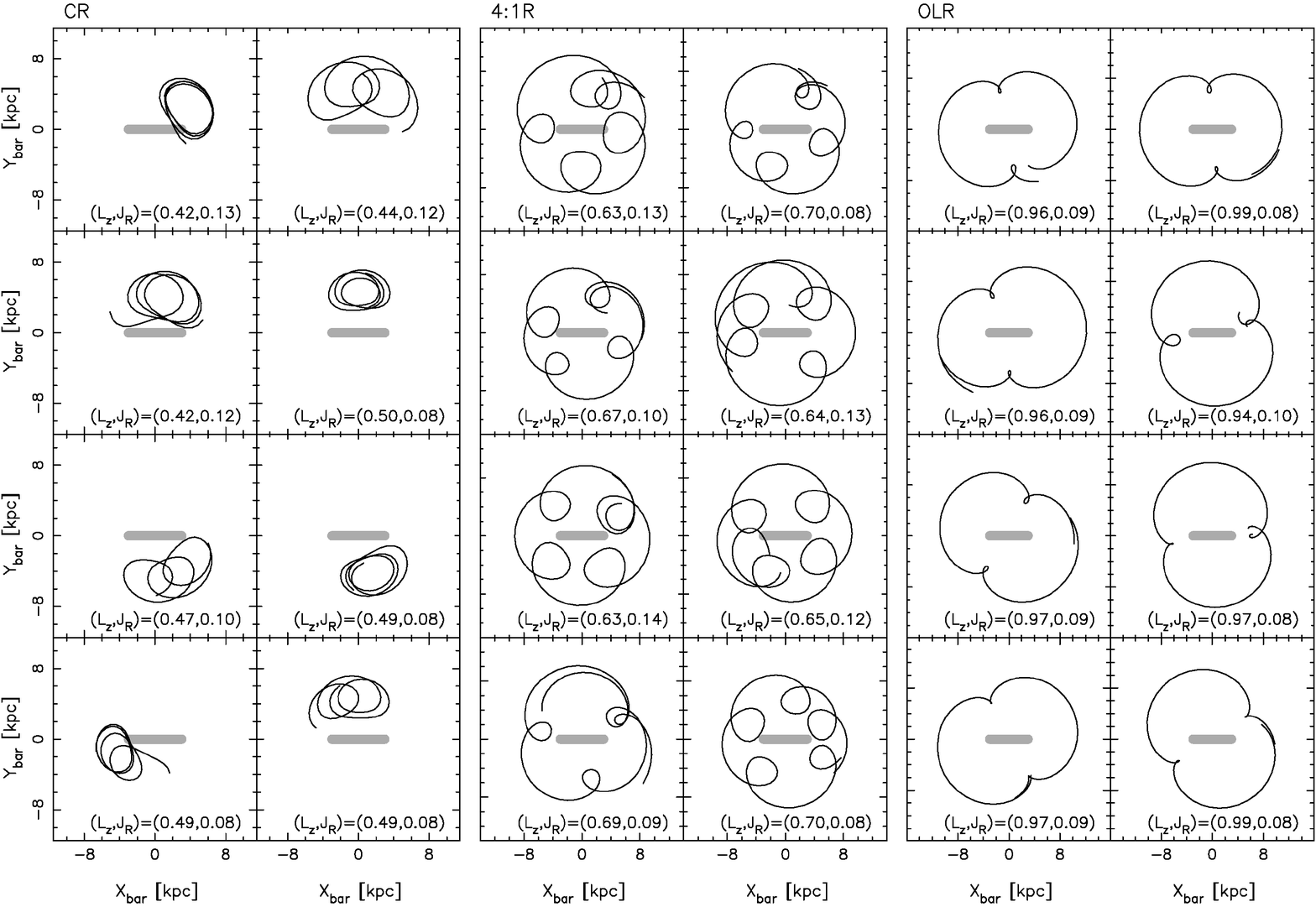}
\caption{Examples of orbits of the particles around CR (left), 4:1R (middle) and OLR (right) with high $J_{\rm R}$ in the bar rotation frame, as the bar is highlighted with the grey horizontal bar. Particle's $L_{\rm z}$ and $J_{\rm R}$ are shown in each panel.}
\label{fig:Rorbits-sim}
\end{figure*}

To make sure that these high $J_{\rm R}$ particles are influenced by the bar resonance, we analyse the orbit of the particles around CR, 4:1R and OLR. Fig.~\ref{fig:Rorbits-sim} shows the orbits of eight randomly selected particles around $J_{\rm R}(L_{\rm z,0})= 0.1$ and around CR, 4:1R and OLR. The orbits are drawn by connecting the position of these particles in the bar rotation frame at the previous outputs. Typical orbits in these resonances are seen for these particles. There are some contaminants from 3:1 and 5:1-like orbits found in 4:1R. However, there are particles with clear 4:1 orbit. Hence, we think that these particles are affected by the bar resonances.

\begin{figure}
 \includegraphics[width=\hsize]{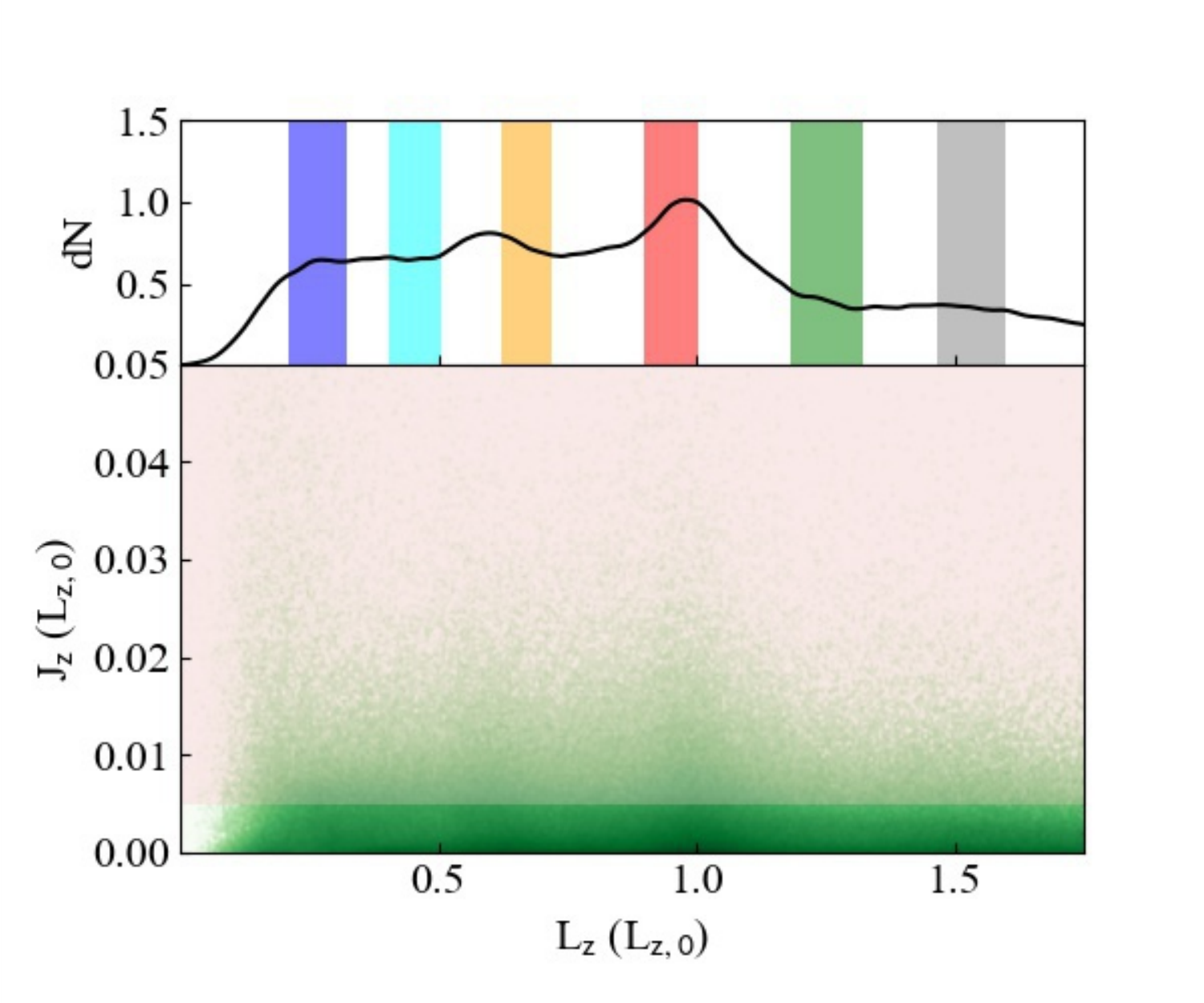}
\caption{The distribution of $L_{\rm z}$ and $J_{\rm z}$ for the star particles of our $N$-body simulation which have $0.07<J_{\rm R}<0.15$, corresponding to the top panel of Fig.~\ref{fig:lzhist-sim}. The upper panel shows the KDE distribution of $L_{\rm z}$ for the star particles with $0.07<J_{\rm R}<0.15$ and
$0.005<J_{\rm z}<0.05$, which are in the pink shaded region in the main panel. The vertical bands highlighted with blue, cyan, orange, red, green and grey respectively indicate the range of i4:1R, CR, 4:1R, OLR, 4:3R and 1:1R for $0.07<J_{\rm R}<0.15$. }
\label{fig:lzjz-sim}
\end{figure}

The rest of the vertical bands highlighted with blue, cyan, green and grey in Fig.~\ref{fig:lzhist-sim} correspond to the i4:1R, CR, 4:3R and 1:1R, respectively. There is no obvious peak around these resonances, except subtle peaks at i4:1R and 1:1R in the upper panel of Fig.~\ref{fig:lzhist-sim}. 
% Still, as more clearly shown in the top and 2nd panel of Fig.~\ref{fig:lzhist-sim} the rate of decrease in the number of star particles with decreasing $L_{\rm z}$ flattens around the CR. In other words, inside the CR there are more stars having higher $J_{\rm R}$. The $L_{\rm z}$ distribution around 4:3R shows similar trend and the decrease rate with increasing $L_{\rm z}$ flattens around 4:3R. However, because it is just next to the strong peak of OLR, it is less clear. 
To further focus on the kinematically hotter stars, we have selected the particles in the upper panel of Fig.~\ref{fig:lzhist-sim}, i.e. particles with $0.07<J_{\rm R}(L_{\rm z,0})<0.15$, and analysed the distribution of $J_{\rm z}$ as a function of $L_{\rm z}$, which is shown in Fig.~\ref{fig:lzjz-sim}. It is interesting to see that there are ridge features toward higher $J_{\rm z}$ around the 4:1R and OLR, which are highlighted with the orange and red bands in the upper panel. This is because more stars are in these resonances and the high $J_{\rm z}$ tail becomes conspicuous at lower $L_{\rm z}$ \citep[see also][]{Trick+Fragkoudi+Hunt+21}. In other words, kinematically hotter stars again show a clearer signal of the stellar number distribution around the resonances. 

Hence, we have further selected high $J_{\rm z}$ star particles, and the upper panel of Fig.~\ref{fig:lzjz-sim} shows the $L_{\rm z}$ KDE distribution of the star particles with $0.005<J_{\rm z}(L_{\rm z,0})<0.05$ (229,099 particles), which are within the region highlighted with the pink shaded region in the main panel. The coloured vertical bands again highlight different resonances. The $L_{\rm z}$ distribution of higher $J_{\rm z}$ star particles show clear peaks around the 4:1 (orange) and OLR (red). There are also some small peaks around i4:1R (blue), 4:3R (green) and 1:1R (grey), though the 4:3R and 1:1R are more tentative. Interestingly, at the CR, we find a small dip or no particular feature in the number of particles. It looks that the CR is unstable for the particles to stay, perhaps because it is where radial migration is efficient \citep[e.g.][]{jsjb02} and many higher order resonances overlap, and/or its overlap with the transient spiral arms \citep{Wu+Pfenniger+Taam16,Hilmi+Minchev+Buck+20} may help to release particles from the CR (Baba et al. in prep.). Investigating mechanisms causing these features is not the aim of this paper. Rather we will use these features seen for kinematically hot star particles, to identify the resonance features in the real Galaxy in the next section. 

% Still, this result indicates that the vertical actions are also affected by the resonances, either by vertically scattering of the particles at the resonances or the higher $J_{\rm z}$ stars trapped at the resonances, when they migrated outward due to the transient arms. 

% To our knowledge, the resonance ridge features in the vertical actions are not reported before. The mechanisms to trigger these features are not known. It could be related to the vertical asymmetry and/or fluctuations of the bar potential in $N$-body simulations \citep[e.g.][]{Hilmi+Minchev+Buck+20}, and further studies are required. It is also interesting to note that the there are significant number of stars having high $J_{\rm z}$ within the CR, which is shown in the flat distribution inside the CR (cyan) in the upper panel of Fig.~\ref{fig:lzjz-sim}. The bar which resides inside the CR could be heating up the star particles vertically and they could be trapped inside the CR. Investigating mechanisms causing these features is not the aim of this paper. Rather we will use these features seen in kinematically hot star particles, to identify the resonances features in the real Galaxy in the next section. 

\section{Bar Resonances Features in {\it Gaia}~EDR3}
\label{sec:gedr3-reso}

Similar to how we analyse our $N$-body/SPH simulation data, we have selected stars in {\it Gaia}~EDR3 and analyse their actions and orbital frequencies. We first select the stars in {\it Gaia}~EDR3 which have radial velocities from {\it Gaia}~RVS \citep[][Seabroke et al.\ in prep.]{Cropper+18}. We then apply quality cuts, selecting stars with renormalised unit weight error, {\tt RUWE} $<1.4$, and $\pi/\sigma_{\pi}>4.0$, where $\pi$ and $\sigma_{\pi}$ are parallax and parallax uncertainty, respectively. We obtain the distance to the stars simply with the inverse of the reported parallax, after applying the zero-point correction suggested by \citet{Lindegren+Bastian+Biermann+21} using the python code provided by the Gaia collaboration\footnote{\url{https://gitlab.com/icc-ub/public/gaiadr3_zeropoint}}. We assume a distance to the Galactic centre from the Sun, $R_0=8.178$~kpc \citep{Gravity+GCdistance19} and the Sun's vertical height from the mid-plane, $z_0=20.8$~pc \citep{Bennett+Bovy19}. We assume the Sun's rotation speed to be $v_{\sun}=248.5$~km~s$^{-1}$ and $v_{z,\sun}=8.5$~km~s$^{-1}$, calculated from the combination of the assumed $R_0$ and the proper motion measurement of Sgr~A$^{*}$ \citep{Reid+Brunthaler20}. We also use the Sun's peculiar motion in the radial direction, $v_{R,\sun}=-12.9$~km~s$^{-1}$ (positive toward the outer disc) and $v_{\sun}-v_{\rm circ}(R_0)=12.32$~km~s$^{-1}$, where $v_{\rm circ}(R_0)$ is the circular velocity at $R_0$. After transforming the data to Galactocentric cylindrical coordinates, we select stars with $4<R<12$~kpc to minimise contamination from spuriously large distances, and $|z|<0.5$~kpc. Using the same method as \citet{Hunt+Johnston+Pettitt+20}, we compute the actions and orbital frequencies of the selected stars using the \texttt{actionAngleStaeckel} \citep{B12-2j} function in \texttt{galpy}, assuming the \texttt{MWPotential2014} potential, which is fit to various observational constraints \citep{jb15}. 
Note that we again normalise orbital frequencies and actions with $\Omega_0=v_{\rm circ}(R_0)/R_0$ and $L_{z,0}=R_0 v_{\rm circ}(R_0)$, respectively, as we did in the previous section.

\begin{figure}
 \includegraphics[width=\hsize]{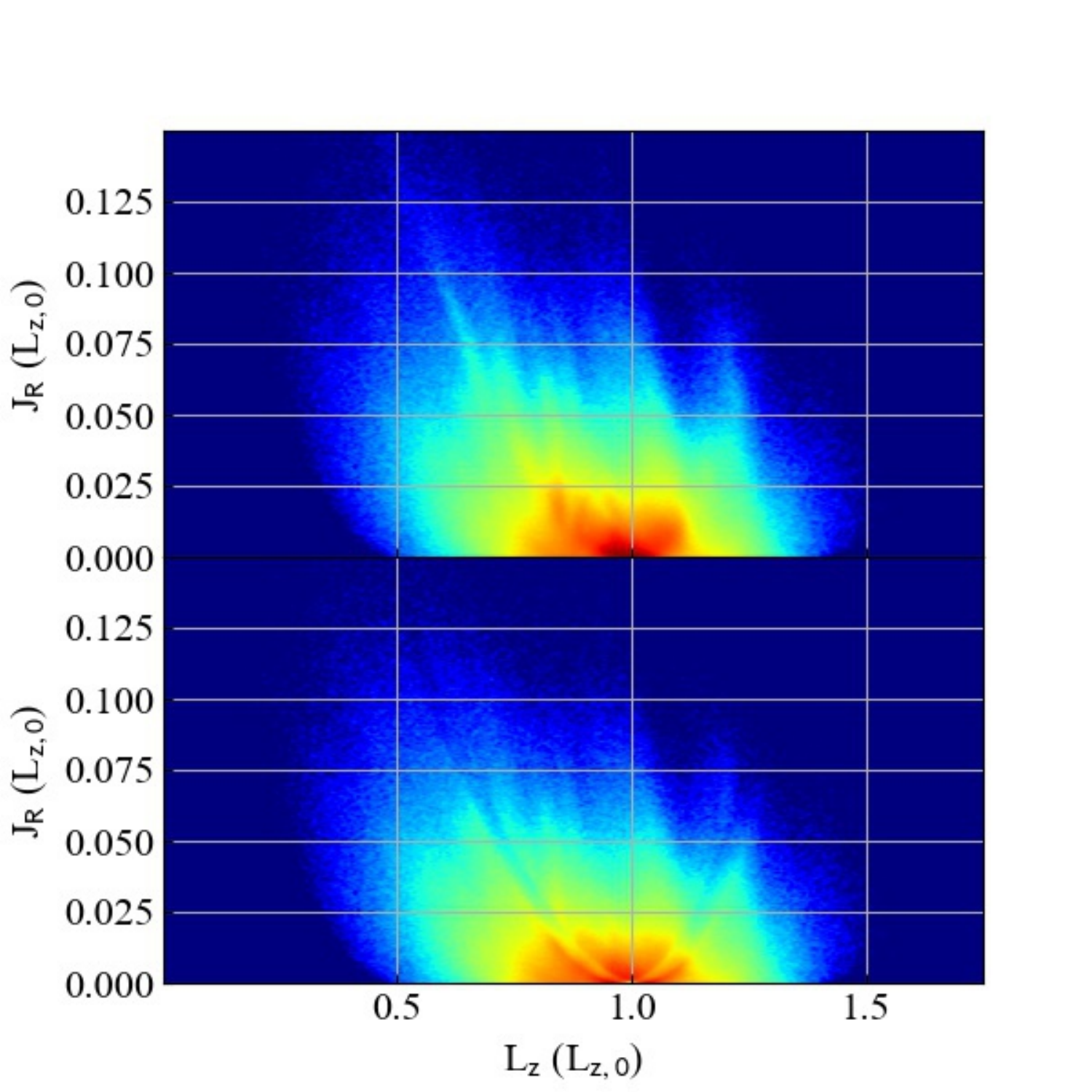}
\caption{Upper panel: The distribution of the angular momentum, $L_{\rm z}$, and radial action, $J_{\rm R}$, for the selected stars in {\it Gaia}~EDR3 without any radius weight. Lower Panel: Same as the upper panel, but also overplotting the stars within distance of 0.1~kpc with white dots.}
\label{fig:lzjr-gedr3-woRw-olrhat}
\end{figure}

The upper panel of Fig.~\ref{fig:lzjr-gedr3-woRw-olrhat} shows the distribution of $L_{\rm z}$ and $J_{\rm R}$ for our selected {\it Gaia}~EDR3 stars. As known from the previous studies with {\it Gaia}~DR2, it is striking to see the various ridge features in this action space. We note that the action space distribution of {\it Gaia}~EDR3 with the RVS data are similar to what is seen in {\it Gaia}~DR2 with the Bayesian distances derived by \citet{Schoenrich+McMillan+Eyer19}. Still, thanks to the superb astrometric accuracy of {\it Gaia}~EDR3, we find tentative new ridge features, as discussed later in more detail.

The upper panel of Fig.~\ref{fig:lzjr-gedr3-woRw-olrhat} shows much finer structures than the result of our $N$-body/SPH simulation in Fig.~\ref{fig:lzjr-sim}, because the {\it Gaia} data are tracing the phase space distribution of stars with much finer resolution, especially in the local volume. However, it also means that the sample is dominated by the stars near the Sun. The lower panel of Fig.~\ref{fig:lzjr-gedr3-woRw-olrhat} shows a similar image to the upper panel, but stars with $|R-R_0|<0.2$~kpc are excluded. We can see a clear parabola feature of an excluded zone centred at $(L_{\rm z}, J_{\rm R})=(1.0L_{\rm z,0},1.0L_{\rm z,0})$, and some of the strong features seen in the upper panel, e.g. a feature extending from $(L_{\rm z}, J_{\rm R})\sim(0.7L_{\rm z,0},0.05L_{\rm z,0})$ to $(0.6L_{\rm z,0},0.1L_{\rm z,0})$, disappear. This demonstrates that these disappeared features are purely due to the dominance of stars close to the Sun.

\begin{figure}
 \includegraphics[width=\hsize]{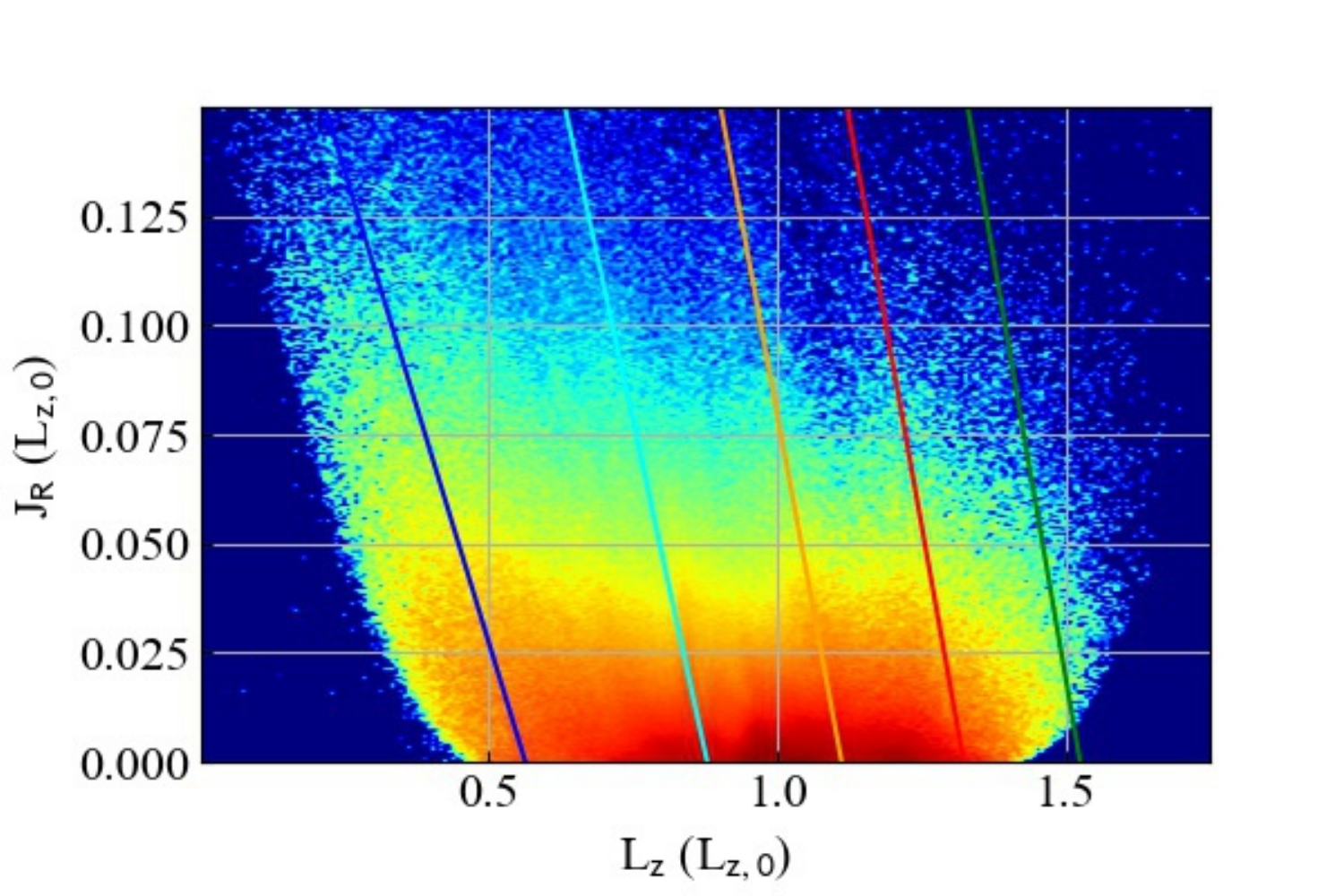}
\caption{The distribution of the angular momentum, $L_{\rm z}$, and radial action, $J_{\rm R}$, for the selected stars in {\it Gaia}~EDR3 when weighting by radius (see the text for more detail). The blue, cyan, orange, red and green solid lines indicate the i4:1R, CR, 4:1R and OLR and 4:3R, respectively, when we assume $\Omega_{\rm bar}\sim1.16\Omega_0$.}
\label{fig:lzjr-gedr3-wRw-olrhat}
\end{figure}

To mitigate this effect, as we did in the previous section, we weight the contribution of stars to this distribution in the action space depending on the Galactocentric radius of the stars, so that the weighted number of stars at different radii becomes constant. The weight for each star is computed with the same method as described in Section~\ref{sec:sim-reso}, but using the 64 radial bins within $4<R<12$~kpc, the different sample radial range in this section. Fig.~\ref{fig:lzjr-gedr3-wRw-olrhat} shows the $L_{\rm z}$ and $J_{\rm R}$ distribution of stars after weighted the stellar contribution to the distribution depending on their Galactocentric radius. Although we can not eliminate the entirety of the selection bias, we at least eliminate the spurious features caused by the overwhelming number of local stars. Hence, in this paper we show the radius weighted results. 

\begin{figure}
 \includegraphics[width=\hsize]{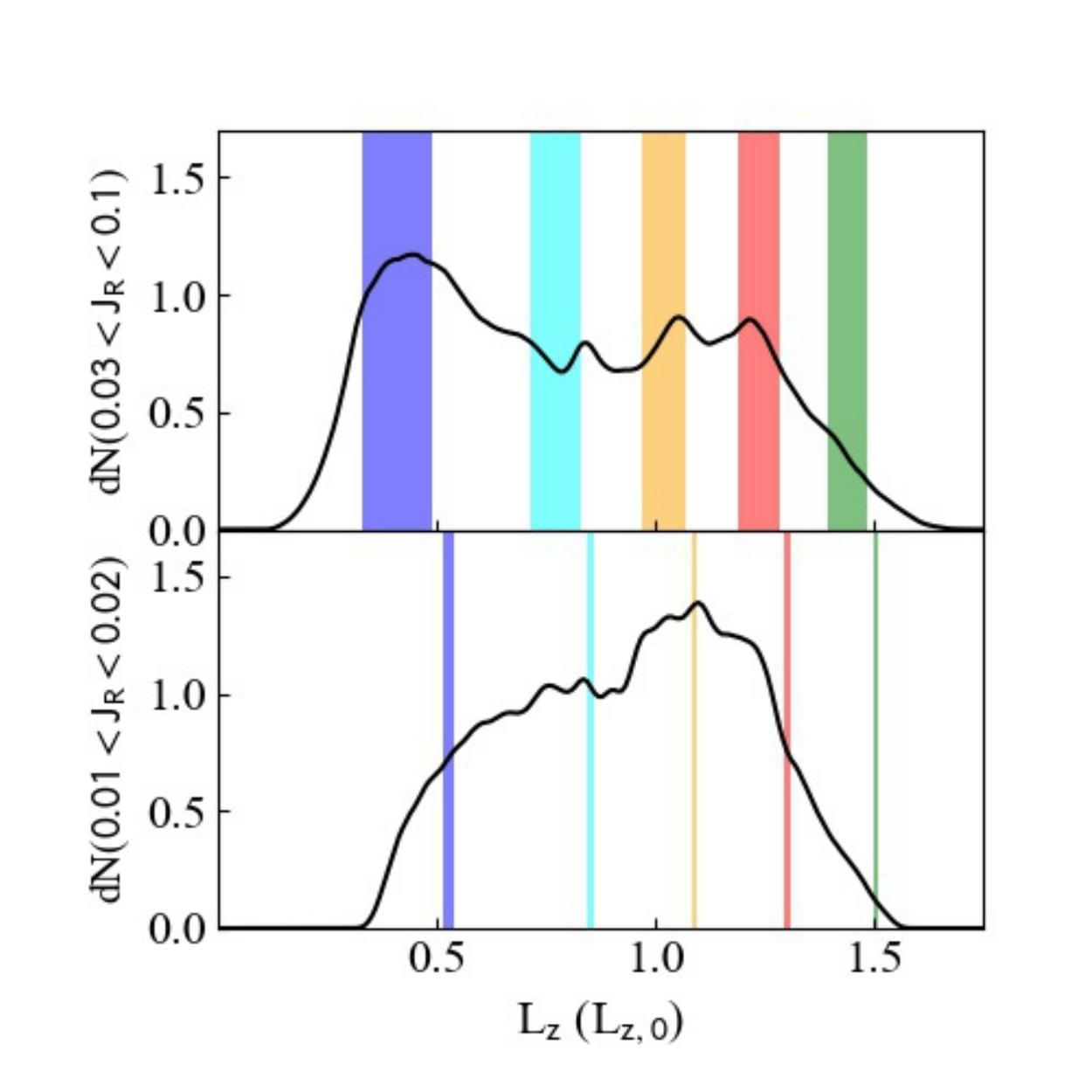}
\caption{The distribution of $L_{\rm z}$ for the {\it Gaia}~EDR3 stars with different ranges of $J_{\rm R}$, $0.03<J_{\rm R}(L_{\rm z,0})<0.1$ (upper) and $0.01<J_{\rm R}(L_{\rm z,0})<0.02$ (lower). The vertical bands highlighted with blue, cyan, orange, red and green indicate the range of i4:1R, CR, 4:1R, OLR and 4:3R, respectively, when we assume $\Omega_{\rm bar}\sim1.16\Omega_0$.}
\label{fig:lzhist-gedr3-wRw-olrhat}
\end{figure}

\begin{figure}
 \includegraphics[width=\hsize]{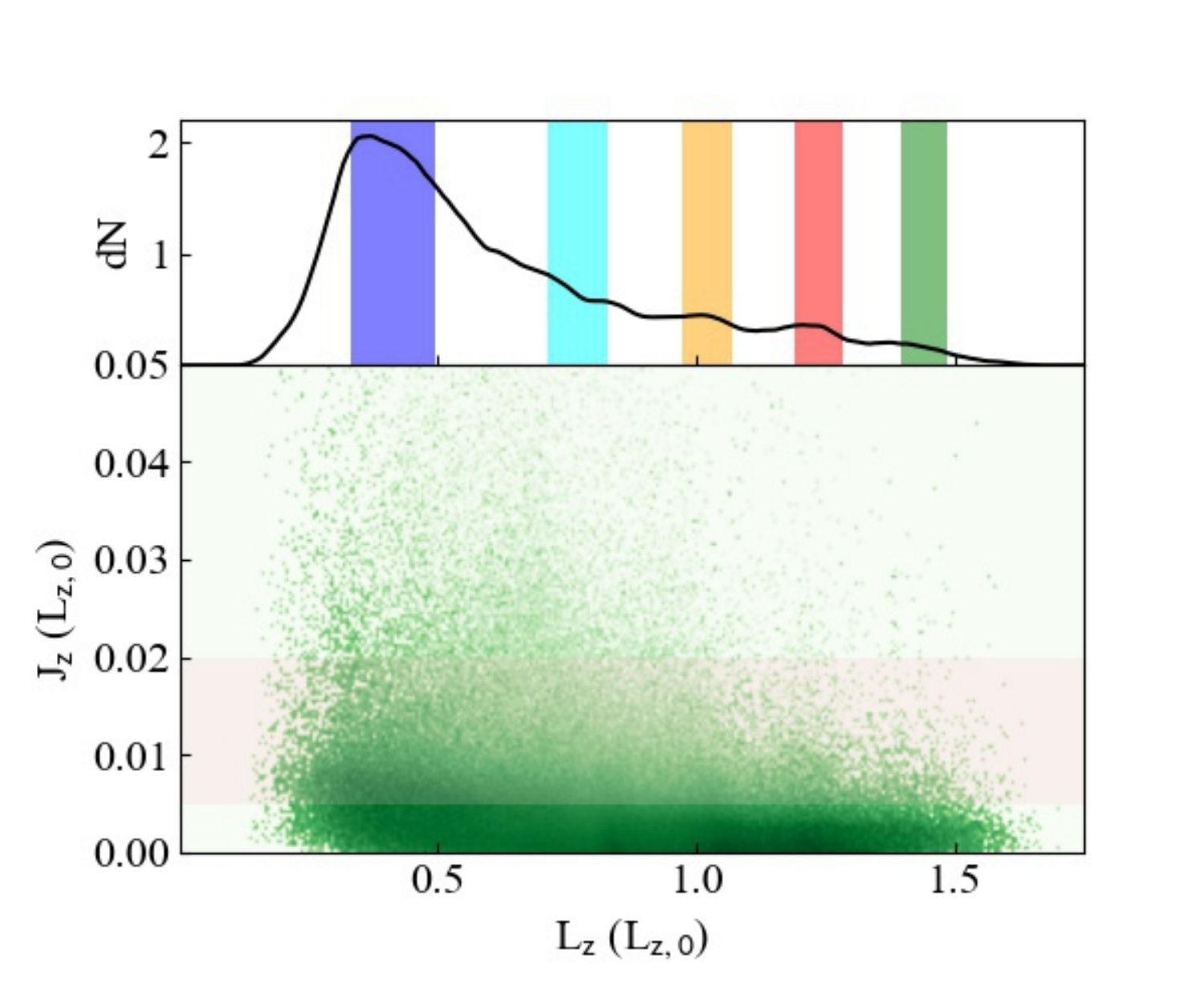}
\caption{The distribution of $L_{\rm z}$ and $J_{\rm z}$ for the stars in {\it Gaia}~EDR3 which have $0.03<J_{\rm R}(L_{\rm z,0})<0.1$, corresponding to the top panel of Fig.~\ref{fig:lzhist-gedr3-wRw-olrhat}. The upper panel shows the KDE distribution of $L_{\rm z}$ for the star particles with $0.03<J_{\rm r}(L_{\rm z,0})<1.0$ and $0.005<J_{\rm z}(L_{\rm z,0})<0.02$, which is highlighted in the pink shaded region in the main panel. The vertical bands highlighted with blue, cyan, orange, red and green respectively indicate the range of the i4:1R, CR, 4:1R, OLR and 4:3R for $0.03<J_{\rm R}(L_{\rm z,0})<0.1$, when we assume $\Omega_{\rm bar}\sim1.16\Omega_0$.}
\label{fig:lzjz-gedr3-wRw-olrhat}
\end{figure}

As seen in our $N$-body/SPH simulations, only a few strong features extend to high $J_{\rm R}$, i.e. $J_{\rm R}>0.03$. As we did in the previous section, Fig.~\ref{fig:lzhist-gedr3-wRw-olrhat} displays the $L_{\rm z}$ KDE distribution of stars with $0.03<J_{\rm R}(L_{\rm z,0})<0.1$ (upper panel) and $0.01<J_{\rm R}(L_{\rm z,0})<0.2$ (lower panel). Note that our selected {\it Gaia}~EDR3 stars show relatively lower action values than our $N$-body/SPH simulation, and therefore we select a lower $J_{\rm R}$ range to pick up the kinematically hot stars. As shown in the previous section using the $N$-body/SPH simulation, we find clearer peaks in the $L_{\rm z}$ distribution for higher $J_{\rm R}$ stars, while colder stars show many smaller peaks, like waves, as discussed in the previous papers with {\it Gaia}~DR2 \citep[e.g.][]{Friske+Schoenrich19,Hunt+Bub+Bovy+19,Trick+Coronado+Rix+19,Trick+Fragkoudi+Hunt+21}. 

Following the strategy of the previous section, we select the stars in the upper panel of Fig.~\ref{fig:lzhist-gedr3-wRw-olrhat}, and analyse the distribution of $L_{\rm z}$ and $J_{\rm z}$ in Fig.~\ref{fig:lzjz-gedr3-wRw-olrhat}. As seen in our $N$-body/SPH simulation (Fig.~\ref{fig:lzjz-sim}), there are several vertically extended ridges for example at $L_{\rm z}\sim L_{\rm z,0}$ and $L_{\rm z}\sim1.2L_{\rm z,0}$. The upper panel shows the $L_{\rm z}$ KDE distribution, when we further restrict the stars with a relatively higher $J_{\rm z}$ range of $0.005<J_{\rm z}(L_{\rm z,0})<0.02$, shaded in pink in the lower panel. We do not select stars with $J_{\rm z}>0.02L_{\rm z,0}$, because these high $J_{\rm z}$ stars are dominated by stars with low angular momentum of $L_{\rm z}\lesssim0.5 L_{\rm z,0}$, which are likely thick disc stars, and which overwhelm the $L_{\rm z}$ distribution, making it difficult to identify smaller peaks. As a result, the $L_{\rm z}$ distribution of our selected stars show several weak, but clear peaks. Thanks to {\it Gaia}~EDR3 where astrometry and radial velocity are precisely measured for a large number of stars, even after this strict selection, there are 163,838 stars which contribute to this distribution. 

Using the prior from our $N$-body/SPH simulation, we consider that these features in high $J_{\rm R}$ and $J_{\rm z}$ stars are due to the bar resonances. Selecting stars with higher actions is analogous to selecting older stars with a higher velocity dispersion, because the actions are correlated with the age of stars in the Milky Way \citep[e.g.][]{Bean+Ness+Bedell18,Ciuca+Kawata+Miglio+20}. As demonstrated with the $N$-body/SPH simulation, we consider that these relatively old (but not as old as the thick disc stars) stars are good tracers to identify the resonances caused by the Galactic bar. A remaining question is which features correspond to which resonances. 

After trying different pattern speeds of the bar and comparing the resonant position with the features in the $L_{\rm z}$ distributions. We find two pattern speeds which equally well explain these features. The first one is a pattern speed of the Galactic bar of $\Omega_{\rm bar}\sim1.16\Omega_0$. Note that exact location of the resonance is sensitive to the shape of the Galactic potential. The pattern speed which we show here is only a rough fit by eye under our assumed potential shape, i.e. {\tt MWPotential2014} from {\tt galpy} \citep{jb15}. The pattern speed value does not mean to be quantitatively accurate, but should only provide a rough estimate. The lines and bands highlighted with blue, cyan, orange, red and green in Figs.~\ref{fig:lzjr-gedr3-wRw-olrhat}, \ref{fig:lzhist-gedr3-wRw-olrhat} and \ref{fig:lzjz-gedr3-wRw-olrhat} correspond to the location of i4:1R, CR, 4:1R, OLR, 4:3R, respectively, when we assume $\Omega_{\rm bar}\sim1.16\Omega_0$. Here, we again use the orbital frequencies of $\Omega_{\rm \phi}$ and $\Omega_{\rm R}$ to identify these resonances, as done in the previous section. 

In the upper panel of Fig.~\ref{fig:lzjz-gedr3-wRw-olrhat} the three weak peaks at $L_{\rm z} \sim L_{\rm z,0}$, $1.2L_{\rm z,0}$ and $1.4L_{\rm z,0}$ nicely aligned with the 4:1R, OLR and 4:3R. A large and broad peak around $L_{\rm z}\sim0.4L_{\rm z,0}$ could potentially be explained by i4:1R. However, as mentioned above, this feature could just be the dominance of the thick disc stars in the inner disc. Interestingly, the CR corresponds to a dip around $L_{\rm z}\sim0.75L_{\rm z,0}$, which is more prominent in the upper panel of Fig.~\ref{fig:lzhist-gedr3-wRw-olrhat}. This is consistent with our $N$-body/SPH simulation, which also shows a subtle dip rather than peak of the $L_{\rm z}$ distribution at the CR. Hence, we regard this as a consistent result with our $N$-body/SPH simulation expectation. In this case, there is no resonance to explain a peak around $L_{\rm z}\sim0.8L_{\rm z,0}$ in top panel of Figs.~\ref{fig:lzhist-gedr3-wRw-olrhat} and \ref{fig:lzjz-gedr3-wRw-olrhat}. However, it can be considered that this peak appears because this is next to the dip of the CR.

If this is the true pattern speed of the Galactic bar, the peak features corresponding to the i4:1R and the 4:3R are newly identified features in the action space, to our knowledge. We believe that the latter one corresponds to the ridge feature found in the radius and rotation velocity distribution in one of {\it Gaia}~EDR3 performance verification papers of \citet{Gaia+Antoja+21}, which they call AC Newridge1. Hence, the peak at $L_{\rm z}\sim1.4$ in the upper panel of Fig.~\ref{fig:lzjz-gedr3-wRw-olrhat} is a confident detection, and this could be the furthest bar resonance feature we have newly identified. As a result, if the bar pattern speed of about $1.16\Omega_0$ is close to the true bar pattern speed, we find the i4:1R, CR, 4:3R, OLR and 4:3R. It is quite remarkable to find i4:1R, which is expected to exist from our simulation in the previous section, and {\it Gaia}~EDR3 may be revealing the resonance inside the Galactic bar. 

\begin{figure}
 \includegraphics[width=\hsize]{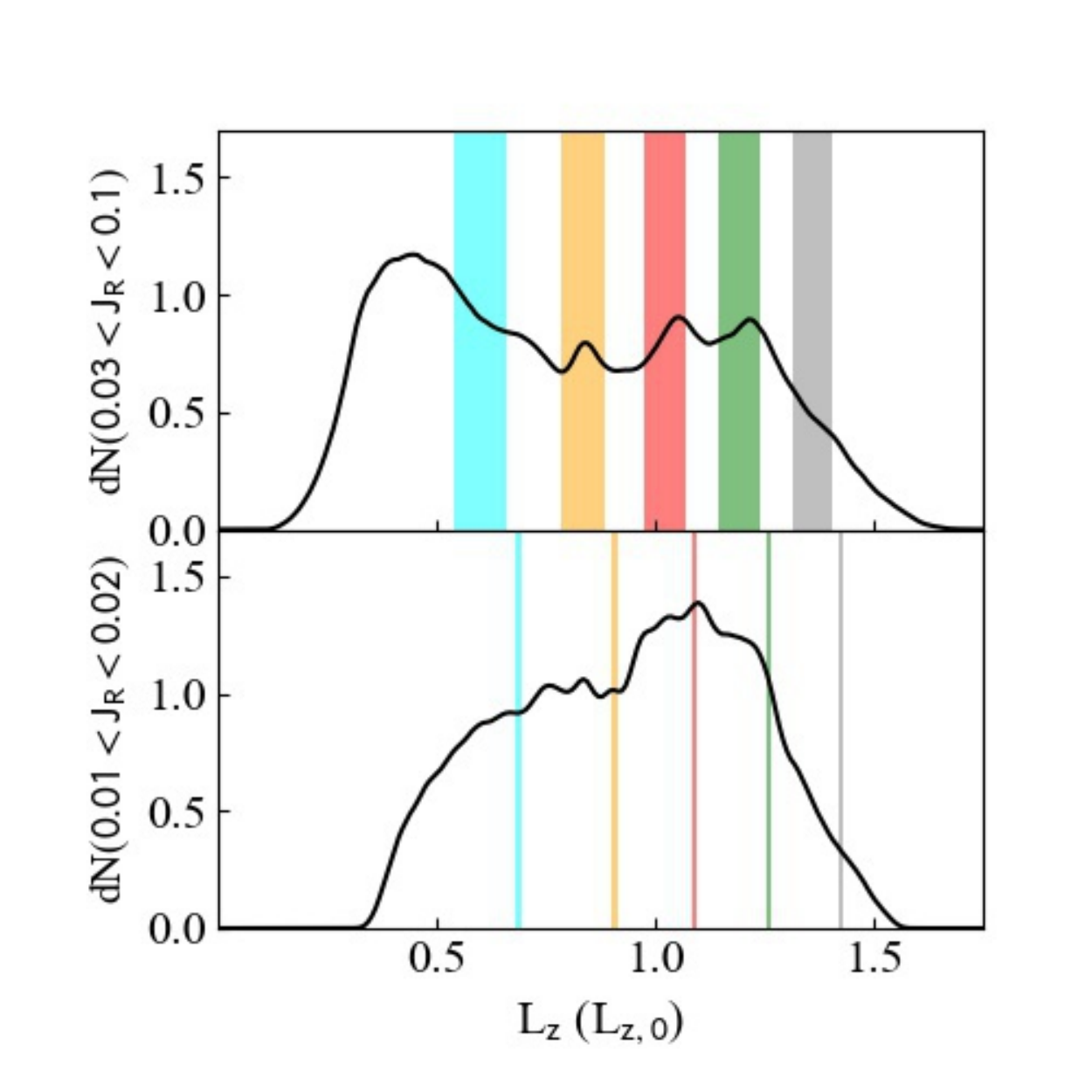}
\caption{The distribution of $L_{\rm z}$ for the {\it Gaia}~EDR3 stars with different ranges of $J_{\rm R}$, $0.03<J_{\rm R}(L_{\rm z,0})<0.1$ (upper) and $0.01<J_{\rm R}(L_{\rm z,0})<0.02$ (lower). The vertical bands highlighted with blue, cyan, orange, red and green indicate the range of CR, 4:1R, OLR, 4:3R and 1:1R, respectively, when we assume $\Omega_{\rm bar}\sim1.45\Omega_0$.}
\label{fig:lzhist-gedr3-wRw-olrsir}
\end{figure}

\begin{figure}
 \includegraphics[width=\hsize]{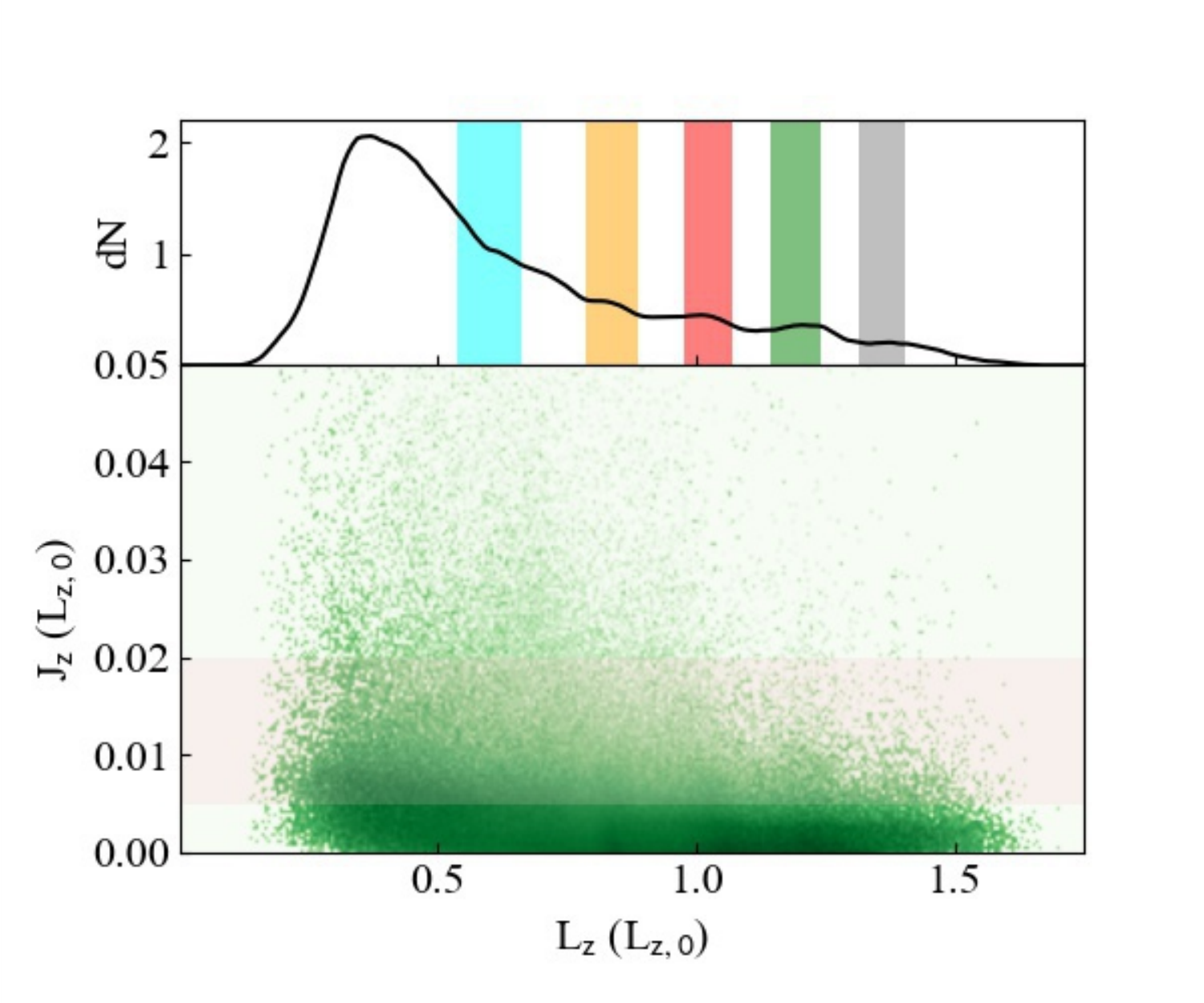}
\caption{The distribution of $L_{\rm z}$ and $J_{\rm z}$ for the stars in {\it Gaia}~EDR3 which have $0.03<J_{\rm R}(L_{\rm z,0})<0.1$, corresponding to the top panel of Fig.~\ref{fig:lzhist-gedr3-wRw-olrsir}. The upper panel shows the distribution of $L_{\rm z}$ for the star particles with $0.03<J_{\rm r}(L_{\rm z,0})<1.0$ and $0.005<J_{\rm z}(L_{\rm z,0})<0.02$, which is highlighted in the pink shaded region in the main panel. The vertical bands highlighted with cyan, orange, red, green and grey respectively indicate the range of CR, 4:1R, OLR, 4:3R and 1:1R for $0.03<J_{\rm R}(L_{\rm z,0})<0.1$, when we assume $\Omega_{\rm bar}\sim1.45\Omega_0$.}
\label{fig:lzjz-gedr3-wRw-olrsir}
\end{figure}

However, we also find that $\Omega_{\rm bar}=1.45\Omega_0$ shows an equally good match to the features in the action distributions. Figs.~\ref{fig:lzhist-gedr3-wRw-olrsir} and \ref{fig:lzjz-gedr3-wRw-olrsir} show the same results as Figs.~\ref{fig:lzhist-gedr3-wRw-olrhat} and \ref{fig:lzjz-gedr3-wRw-olrhat}, but overlaid with the position of the CR, 4:1R, OLR, 4:3R and 1:1R, when a bar pattern speed of $1.45\Omega_0$ is assumed. In this case, we consider the peak at $L_{\rm z}\sim0.8L_{\rm z,0}$ in the top panels of Figs.~\ref{fig:lzhist-gedr3-wRw-olrsir} and \ref{fig:lzjz-gedr3-wRw-olrsir} to be due to the 4:3R. Again, the CR corresponds to the dip around $L_{\rm z}\sim0.6L_{\rm z,0}$. However, this dip is not as clear as the one which we associate with the CR, when assuming $\Omega_{\rm bar}=1.16\Omega_0$. With this pattern speed, the furthest resonance is associated to the 1:1R, which is also expected to be visible from the prediction of our $N$-body/SPH simulation in the previous section. As a result, if the bar pattern speeds is $1.45\Omega_0$, our identified features correspond to the CR, 4:1R, OLR, 4:3R and 1:1R. It is also remarkable to identify the 1:1R.

\begin{figure}
 \includegraphics[width=\hsize]{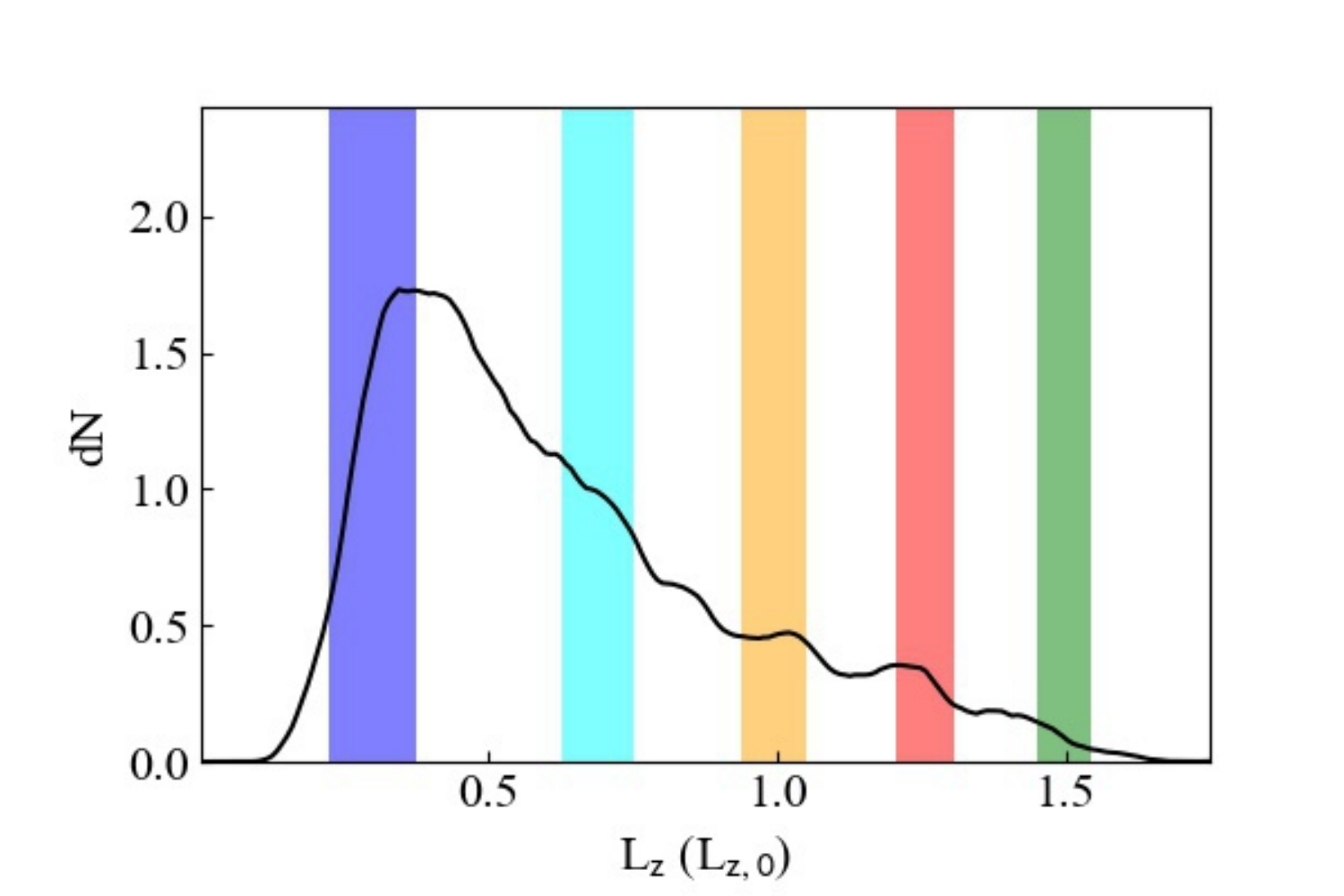}
\caption{The KDE distribution of $L_{\rm z}$ for the star particles with $0.03<J_{\rm r}(L_{\rm z,0})<1.0$ and $0.005<J_{\rm z}(L_{\rm z,0})<0.02$, when Galactic potential of \citet{McMillan17} is employed. The vertical bands highlighted with blue, cyan, orange, red and green respectively indicate the range of the i4:1R, CR, 4:1R, OLR and 4:3R for $0.03<J_{\rm R}(L_{\rm z,0})<0.1$, when we assume $\Omega_{\rm bar}\sim 35.5$~km~s$^{-1}$~kpc$^{-1}$ to align the OLR to the Hat local kinematic feature.}
\label{fig:lzjz-gedr3-wRw-olrhat-M17}
\end{figure}

\begin{figure}
 \includegraphics[width=\hsize]{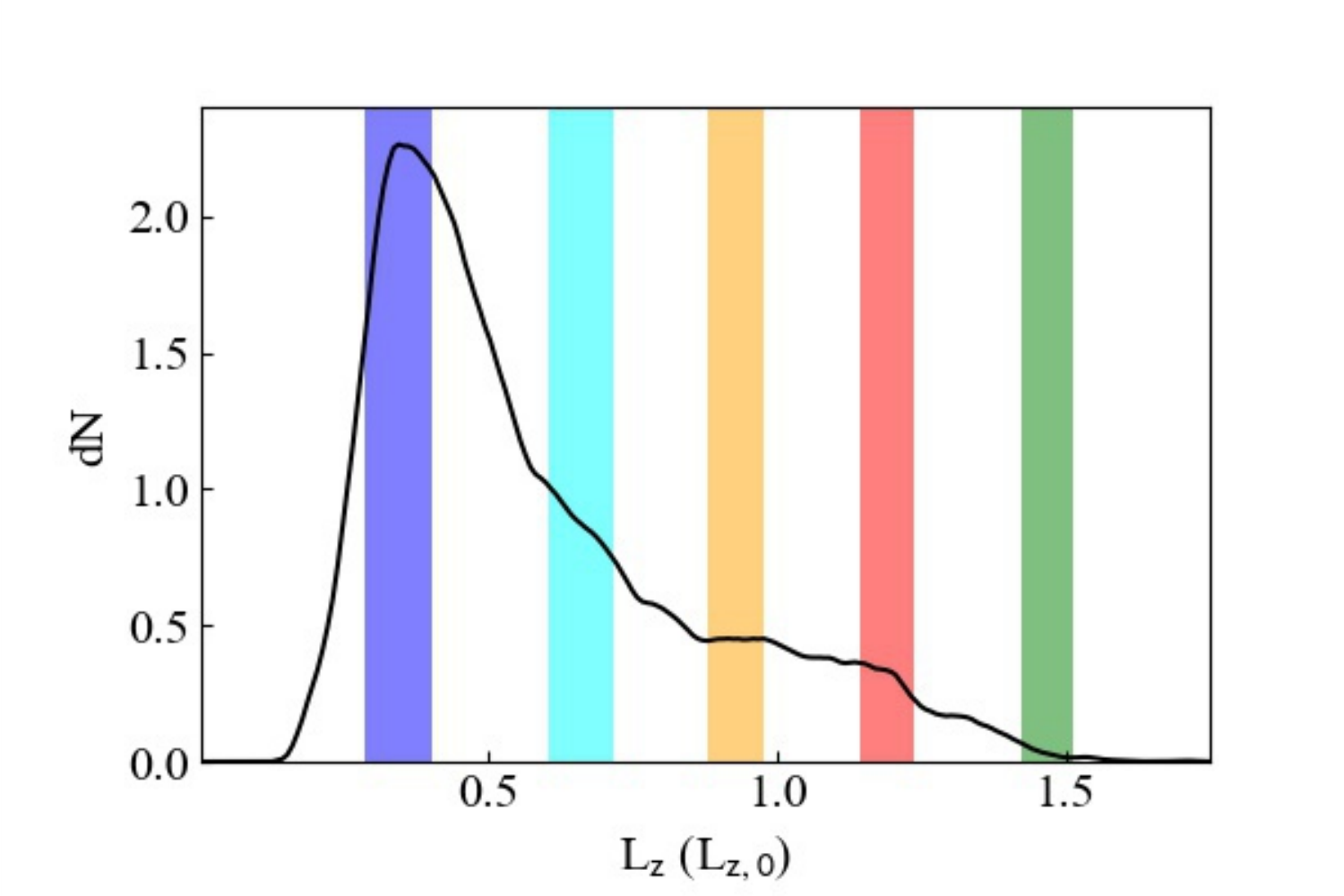}
\caption{The KDE distribution of $L_{\rm z}$ for the star particles with $0.03<J_{\rm r}(L_{\rm z,0})<1.0$ and $0.005<J_{\rm z}(L_{\rm z,0})<0.02$, when Galactic potential of \citet{Irrgang+13} is employed. The vertical bands highlighted with blue, cyan, orange, red and green respectively indicate the range of the i4:1R, CR, 4:1R, OLR and 4:3R for $0.03<J_{\rm R}(L_{\rm z,0})<0.1$, when we assume $\Omega_{\rm bar}\sim 35.5$~km~s$^{-1}$~kpc$^{-1}$ to align the OLR to the Hat local kinematic feature.}
\label{fig:lzjz-gedr3-wRw-olrhat-I13}
\end{figure}

\section{Summary and Discussion}
\label{sec:sum}

Using an $N$-body/SPH simulation of an isolated barred disc galaxy, we demonstrate that the resonances of the bar induce more prominent features in the action space distribution for the kinematically hot star particles, i.e. particles with relatively high actions, than the kinematically colder star particles. This is because kinematically colder stars are more affected by the weaker but local non-axisymmetric strucures, such as transient spiral arms, than the kinematically hotter stars. Using this as a working assumption, we analyse the action distribution for the kinematically hotter stars identified in the recently provided {\it Gaia}~EDR3 data with the radial velocities from {\it Gaia}~RVS. After computing the actions of these stars, we find several features in the angular momentum, $L_{\rm z}$, distribution for kinematically hot, relatively high $J_{\rm R}$ and high $J_{\rm z}$ stars. Due to the improved astrometry in {\it Gaia}~EDR3, we find new ridge features extending from around $L_{\rm z}=0.6L_{\rm z,0}$ and $L_{\rm z}=1.5L_{\rm z,0}$ at $J_{\rm R}=0$, although the features are tentative and close to the edge of the data. 

Assuming these features correspond to the bar resonances as seen in our $N$-body/SPH simulation, we find that bar pattern speeds of $\Omega_{\rm bar}=1.16\Omega_0$ and $\Omega_{\rm bar}=1.45\Omega_0$ both explain all these features well. With our assumed $R_0$ and $v_{\rm circ}(R_0)$, these pattern speeds correspond to 33.6~km~s$^{-1}$~kpc$^{-1}$ and 42~km~s$^{-1}$~kpc$^{-1}$, respectively. When we adopt $\Omega_{\rm bar}=1.16\Omega_0$, the features correspond to the i4:1R, CR, 4:1R, OLR and 4:3R. When we adopt $\Omega_{\rm bar}=1.45\Omega_0$, the features can be explained by the CR, 4:1R, OLR, 4:3R and 1:1R. To our knowledge, these many resonance features have never been revealed in the Milky Way before. This demonstrates the power of the {\it Gaia}~EDR3 data.

Interestingly, in both cases, the CR is identified as a dip rather than peak in the $L_{\rm z}$ distribution. We find a similar deficit of the particles at the CR in the $L_{\rm z}$ distribution of the $N$-body/SPH simulation, and hence we regard this dip (or no peak) as a consistent feature of the CR. The mechanism responsible for the deficit of stars at the CR is not clear, and it requires more theoretical studies with both a bar and transient spiral arms, which our $N$-body/SPH simulation has, and which are expected to impact the stellar motion. 

The bar pattern speed of $1.16 \Omega_0$ is consistent with what was suggested by \citet{Perez-Villegas+17,Monari+Famaey+Siebert+Wegg+Gerhard19} to explain the Hercules stream with the CR and also consistent with the current pattern speed suggested by \citet{Chiba+Friske+Schoenrich20,Chiba+Schoenrich21} with a slowing bar. This is also roughly consistent with the recently measured pattern speed of $\Omega_{\rm bar}=35.4\pm0.9$~km~s$^{-1}$~kpc$^{-1}$ from dynamical modelling of the proper motion of the stars in the bar region \citep{Clarke+Gerhard21}. This slow bar pattern speed puts the OLR on the so-called "Hat" phase space velocity feature identified in the Solar neighbourhood radial and rotational velocity distribution \citep{Hunt+17,Hunt+Bovy18}. Although we consider that the local velocity fields can be disturbed by the transient spiral arms \citep{Hunt+Hong+Bovy+18} and are not necessarily reliable indicators of the resonance features, \citet{Trick+Fragkoudi+Hunt+21} discussed that at the OLR the radial motion of the stars flips from outward motion (inside the OLR) to inward motion (outside the OLR), and the Hat has this characteristic flip expected at the OLR.

The other pattern speed of $1.45\Omega_0$ is consistent with what is inferred from the kinematics of the stars in the Galactic bar, which are converging to around $1.45\Omega_0$ as suggested by \citet{Sanders+Smith+Evans19,Bovy+Leung+Hunt+19}. \citet{Trick+Fragkoudi+Hunt+21} discussed that this pattern speed places the OLR on the Sirius moving group, where they do not see the outward to inward motion flip. \citet{Ramos+18} discuss that the Sirius moving group is unlikely to be induced by the resonance, because their kinematic feature is more consistent with the constant kinetic energy rather than the constant angular momentum. However, we consider that the velocity fields could be distorted by perturbers, such as transient spiral arms or the impact of Sagittarius, and we do not require these kinematic features
%this motion flip 
to occur at the resonances in this paper. Hence, we also consider this to be an acceptable pattern speed. In this case, the strong Hat ridge feature is explained with the 4:3R. Our conclusion of the two potential bar pattern speeds whose OLR corresponds to Hat or Sirius is consistent with the conclusion from the phase-angle analysis in \citet{Trick21}.

We note that as mentioned in Section~\ref{sec:gedr3-reso}, the computed values of actions depend on the shape of the Galactic potential in the radial range of the orbits of our sampled stars. The results of this paper are under the assumption of the Galactic potential of {\tt MWPotential2014} in {\tt galpy}. To note the dependence on this assumption, we briefly discuss the results when we adopt the different shapes of the Galactic potential. To this end, we compute the actions of our sampled stars using {\tt McMillan17} \citep{McMillan17} and {\tt Irrgang13III} \citep[Model III of][]{Irrgang+13} Galactic potentials in {\tt galpy}. We then applied the same selection of kinematically hot stars as the top panel of Fig.~\ref{fig:lzjz-gedr3-wRw-olrhat}, i.e. $0.03<J_{\rm r}(L_{\rm z,0})<1.0$ and $0.005<J_{\rm z}(L_{\rm z,0})<0.02$. The angular momentum distribution of these stars are displayed in Figs.~\ref{fig:lzjz-gedr3-wRw-olrhat-M17} and \ref{fig:lzjz-gedr3-wRw-olrhat-I13}. Here, we normalise the actions and angular momentum using $(R_0, v_{\rm circ}(R_0))=$ (8.21~kpc, 233.1~km~s$^{-1}$) and (8.33~kpc, 239.7~km~s$^{-1}$) for {\tt McMillan17} and {\tt Irrgang13III} potentials, respectively. The vertical bands indicating the resonances are computed with the bar pattern speeds of $\Omega_{\rm bar}=35.5$~km~s$^{-1}$~kpc$^{-1}$~$\sim1.25\Omega_0$ ({\tt McMillan17}) and 39~km~s$^{-1}$~kpc$^{-1}$~$\sim1.36\Omega_0$ ({\tt Irrgang13III}), which are chosen to match the OLR with the Hat local kinematic feature as did in Fig.~\ref{fig:lzjz-gedr3-wRw-olrhat}. 

These results indicate that the required pattern speed of the bar to locate the OLR to the Hat feature depends on the shape of the Galactic potential. Also, the intervals of the resonance locations in $L_{\rm z}$ are sensitive to the assumed Galactic potential, and the locations of i4:1R, CR and 4:3R with respect to the dips and peaks of the $L_{\rm z}$ distribution are different from Fig.~\ref{fig:lzjz-gedr3-wRw-olrhat} with {\tt MWPotential2014}. Both {\tt McMillan17} and {\tt Irrgang13III} place the CR around the peak of the $L_{\rm z}$ distribution, next to (lower side of) the dip which coincides with the CR in {\tt MWPotential2014} in
Fig.~\ref{fig:lzjz-gedr3-wRw-olrhat}. Also, the locations of i4:1R and 4:3R are not aligned well with the peaks, compared to the results in Fig.~\ref{fig:lzjz-gedr3-wRw-olrhat}. Although this is beyond the scope of this paper, this may indicate that {\tt MWPotential2014} is a preferable shape of Galactic potential, because of better matching of the resonances with the peaks and dips of the $L_{\rm z}$ distribution. In other words, alignment of the resonances with the $L_{\rm z}$ distribution of kinematically hot stars could constrain the shape of the Galactic potential, if the features identified in this papers are truly induced by the resonances of the Galactic bar, as expected from our N-body simulation.

Although we tried to correct the bias due to the overwhelming number of local stars by weighting the contribution of stars in the analysis depending on their Galactocentric radius, we should still be careful with the observational selection bias which could influence our conclusions. Also, we must always remind ourselves that $N$-body/SPH simulations can still be far from a true representation of the real Milky Way. Our simulation formed the bar at about 5.5~Gyr before the snapshot we used for this study. Also, the pattern speed of the bar does not change significantly since its formation, because of the rigid dark matter halo we used. Hence, the disc star particles were subject to the same resonance locations for a long time. The effect on the action distribution of the disc stars could be significantly different, if the bar of the Milky Way formed recently, or if the pattern speed of the bar was slowing down \citep{Chiba+Friske+Schoenrich20, Chiba+Schoenrich21}. Also, if the Galactic disc recently experienced bar-buckling \citep[e.g.][]{Khoperskov+DiMatteo+Gerhard+19} and/or perturbations from the satellite galaxies, such as the Sagittarius dwarf \citep[e.g.][]{LMJG19}, their effects could be significant enough to erase the resonance features. Further comparison between the observational data and the theoretical models taking into account all these potential effects would be necessary to recover the true Galactic bar pattern speed confidently, and ultimately understand the nature of the Galactic bar and its impact on the Galactic disc evolution. To this end, obtaining precise proper motions inside the Galactic bar will be crucial \citep[e.g.][]{Baba+Kawata20}. The upcoming near-infrared astrometry mission, {\it Japan Astrometry Satellite Mission for INfrared Exploration} \citep[{\it  JASMINE};][]{Gouda+20}\footnote{\url{http://jasmine.nao.ac.jp/index-en.html}}, will be invaluable in providing proper motion of stars between the Sun and the Galactic centre. 

\section*{Data Availability}

The data underlying this article will be shared on reasonable request to the corresponding author.

 \section*{Acknowledgments}
We thank anonymous referee for his/her helpful suggestions that have improved the manuscript significantly.
DK, RS, IC, JF, MC and GS acknowledge the support of the UK's Science \& Technology Facilities Council (STFC Grant ST/S000216/1 and ST/S000984/1). JB acknowledge the supports by the Japan Society for the Promotion of Science (JSPS) KAKENHI grant Nos. 18K03711, 18H01248, 19H01933, 21K03633 and 21H00054. JH is supported by a Flatiron Research Fellowship at the Flatiron institute, which is supported by the Simons Foundation. RS is supported by a Royal Society University Research Fellowship. JF is supported by a UCL Graduate Research Scholarship, the Ev. Studienwerk Villigst and the Max-Weber-Programm. Calculations and analyses of our simulated galaxy were carried out on Cray XC50 (ATERUI-II) and computers at Center for Computational Astrophysics, National Astronomical Observatory of Japan (CfCA/NAOJ). This work was inspired from our numerical simulation studies used the UCL facility Grace and the DiRAC Data Analytic system at the University of Cambridge, operated by the University of Cambridge High Performance Computing Service on behalf of the STFC DiRAC HPC Facility (\url{www.dirac.ac.uk}). This equipment was funded by BIS National E-infrastructure capital grant (ST/K001590/1), STFC capital grants ST/H008861/1 and ST/H00887X/1, and STFC DiRAC Operations grant ST/K00333X/1. DiRAC is part of the National E-Infrastructure. This work has made use of data from the European Space Agency (ESA) mission {\it Gaia} (\url{https://www.cosmos.esa.int/gaia}), processed by the {\it Gaia} Data Processing and Analysis Consortium (DPAC, \url{https://www.cosmos.esa.int/web/gaia/dpac/consortium}). Funding for the DPAC has been provided by national institutions, in particular the institutions participating in the {\it Gaia} Multilateral Agreement.   
%%%%%%%%%%%%%%%%%%%%%%%%%%%%%%%%%%%%%%%%%%%%%%%%%%



\bibliographystyle{mnras}
\bibliography{./dkref} % if your bibtex file is called example.bib

\begin{thebibliography}{}
\makeatletter
\relax
\def\mn@urlcharsother{\let\do\@makeother \do\$\do\&\do\#\do\^\do\_\do\%\do\~}
\def\mn@doi{\begingroup\mn@urlcharsother \@ifnextchar [ {\mn@doi@}
  {\mn@doi@[]}}
\def\mn@doi@[#1]#2{\def\@tempa{#1}\ifx\@tempa\@empty \href
  {http://dx.doi.org/#2} {doi:#2}\else \href {http://dx.doi.org/#2} {#1}\fi
  \endgroup}
\def\mn@eprint#1#2{\mn@eprint@#1:#2::\@nil}
\def\mn@eprint@arXiv#1{\href {http://arxiv.org/abs/#1} {{\tt arXiv:#1}}}
\def\mn@eprint@dblp#1{\href {http://dblp.uni-trier.de/rec/bibtex/#1.xml}
  {dblp:#1}}
\def\mn@eprint@#1:#2:#3:#4\@nil{\def\@tempa {#1}\def\@tempb {#2}\def\@tempc
  {#3}\ifx \@tempc \@empty \let \@tempc \@tempb \let \@tempb \@tempa \fi \ifx
  \@tempb \@empty \def\@tempb {arXiv}\fi \@ifundefined
  {mn@eprint@\@tempb}{\@tempb:\@tempc}{\expandafter \expandafter \csname
  mn@eprint@\@tempb\endcsname \expandafter{\@tempc}}}

\bibitem[\protect\citeauthoryear{{Antoja} et~al.,}{{Antoja}
  et~al.}{2018}]{Antoja+18}
{Antoja} T.,  et~al., 2018, \mn@doi [\nat] {10.1038/s41586-018-0510-7}, \href
  {https://ui.adsabs.harvard.edu/abs/2018Natur.561..360A} {561, 360}

\bibitem[\protect\citeauthoryear{{Asano}, {Fujii}, {Baba}, {B{\'e}dorf},
  {Sellentin}  \& {Portegies Zwart}}{{Asano} et~al.}{2020}]{Asano+20}
{Asano} T.,  {Fujii} M.~S.,  {Baba} J.,  {B{\'e}dorf} J.,  {Sellentin} E.,
  {Portegies Zwart} S.,  2020, \mn@doi [\mnras] {10.1093/mnras/staa2849}, \href
  {https://ui.adsabs.harvard.edu/abs/2020MNRAS.499.2416A} {499, 2416}

\bibitem[\protect\citeauthoryear{{Baba} \& {Kawata}}{{Baba} \&
  {Kawata}}{2020}]{Baba+Kawata20}
{Baba} J.,  {Kawata} D.,  2020, \mn@doi [\mnras] {10.1093/mnras/staa140}, \href
  {https://ui.adsabs.harvard.edu/abs/2020MNRAS.492.4500B} {492, 4500}

\bibitem[\protect\citeauthoryear{{Baba}, {Saitoh}  \& {Wada}}{{Baba}
  et~al.}{2013}]{bsw13}
{Baba} J.,  {Saitoh} T.~R.,   {Wada} K.,  2013, \mn@doi [\apj]
  {10.1088/0004-637X/763/1/46}, \href
  {http://adsabs.harvard.edu/abs/2013ApJ...763...46B} {763, 46}

\bibitem[\protect\citeauthoryear{{Baba}, {Morokuma-Matsui}  \& {Saitoh}}{{Baba}
  et~al.}{2017}]{Baba+2017}
{Baba} J.,  {Morokuma-Matsui} K.,   {Saitoh} T.~R.,  2017, \mn@doi [\mnras]
  {10.1093/mnras/stw2378}, \href {http://ads.nao.ac.jp/abs/2017MNRAS.464..246B}
  {464, 246}

\bibitem[\protect\citeauthoryear{{Baba}, {Kawata}  \& {Sch{\"o}nrich}}{{Baba}
  et~al.}{2021}]{Baba+Kawata+Schoenrich21}
{Baba} J.,  {Kawata} D.,   {Sch{\"o}nrich} R.,  2021, arXiv e-prints, \href
  {https://ui.adsabs.harvard.edu/abs/2021arXiv210409526B} {p. arXiv:2104.09526}

\bibitem[\protect\citeauthoryear{{Beane}, {Ness}  \& {Bedell}}{{Beane}
  et~al.}{2018}]{Bean+Ness+Bedell18}
{Beane} A.,  {Ness} M.~K.,   {Bedell} M.,  2018, \mn@doi [\apj]
  {10.3847/1538-4357/aae07f}, \href
  {https://ui.adsabs.harvard.edu/abs/2018ApJ...867...31B} {867, 31}

\bibitem[\protect\citeauthoryear{{Bennett} \& {Bovy}}{{Bennett} \&
  {Bovy}}{2019}]{Bennett+Bovy19}
{Bennett} M.,  {Bovy} J.,  2019, \mn@doi [\mnras] {10.1093/mnras/sty2813},
  \href {https://ui.adsabs.harvard.edu/abs/2019MNRAS.482.1417B} {482, 1417}

\bibitem[\protect\citeauthoryear{{Binney}}{{Binney}}{2012}]{B12-2j}
{Binney} J.,  2012, \mn@doi [\mnras] {10.1111/j.1365-2966.2012.21757.x}, \href
  {https://ui.adsabs.harvard.edu/abs/2012MNRAS.426.1324B} {426, 1324}

\bibitem[\protect\citeauthoryear{{Binney}}{{Binney}}{2018}]{Binney18a}
{Binney} J.,  2018, \mn@doi [\mnras] {10.1093/mnras/stx2835}, \href
  {https://ui.adsabs.harvard.edu/abs/2018MNRAS.474.2706B} {474, 2706}

\bibitem[\protect\citeauthoryear{{Binney} \& {Tremaine}}{{Binney} \&
  {Tremaine}}{2008}]{Binney+Tremaine08}
{Binney} J.,  {Tremaine} S.,  2008, {Galactic Dynamics: Second Edition}.
Princeton University Press

\bibitem[\protect\citeauthoryear{{Bland-Hawthorn} \&
  {Gerhard}}{{Bland-Hawthorn} \& {Gerhard}}{2016}]{bhg16}
{Bland-Hawthorn} J.,  {Gerhard} O.,  2016, \mn@doi [\araa]
  {10.1146/annurev-astro-081915-023441}, \href
  {http://adsabs.harvard.edu/abs/2016ARA%26A..54..529B} {54, 529}

\bibitem[\protect\citeauthoryear{{Bovy}}{{Bovy}}{2015}]{jb15}
{Bovy} J.,  2015, \mn@doi [\apjs] {10.1088/0067-0049/216/2/29}, \href
  {http://adsabs.harvard.edu/abs/2015ApJS..216...29B} {216, 29}

\bibitem[\protect\citeauthoryear{{Bovy}, {Leung}, {Hunt}, {Mackereth},
  {Garc{\'\i}a-Hern{\'a}ndez}  \& {Roman-Lopes}}{{Bovy}
  et~al.}{2019}]{Bovy+Leung+Hunt+19}
{Bovy} J.,  {Leung} H.~W.,  {Hunt} J. A.~S.,  {Mackereth} J.~T.,
  {Garc{\'\i}a-Hern{\'a}ndez} D.~A.,   {Roman-Lopes} A.,  2019, \mn@doi
  [\mnras] {10.1093/mnras/stz2891}, \href
  {https://ui.adsabs.harvard.edu/abs/2019MNRAS.490.4740B} {490, 4740}

\bibitem[\protect\citeauthoryear{{Chiba} \& {Sch{\"o}nrich}}{{Chiba} \&
  {Sch{\"o}nrich}}{2021}]{Chiba+Schoenrich21}
{Chiba} R.,  {Sch{\"o}nrich} R.,  2021, \mn@doi [\mnras]
  {10.1093/mnras/stab1094}, \href
  {https://ui.adsabs.harvard.edu/abs/2021MNRAS.505.2412C} {505, 2412}

\bibitem[\protect\citeauthoryear{{Chiba}, {Friske}  \& {Sch{\"o}nrich}}{{Chiba}
  et~al.}{2021}]{Chiba+Friske+Schoenrich20}
{Chiba} R.,  {Friske} J. K.~S.,   {Sch{\"o}nrich} R.,  2021, \mn@doi [\mnras]
  {10.1093/mnras/staa3585}, \href
  {https://ui.adsabs.harvard.edu/abs/2020MNRAS.tmp.3375C} {500, 4710}

\bibitem[\protect\citeauthoryear{{Ciuc{\u{a}}}, {Kawata}, {Miglio}, {Davies}
  \& {Grand}}{{Ciuc{\u{a}}} et~al.}{2021}]{Ciuca+Kawata+Miglio+20}
{Ciuc{\u{a}}} I.,  {Kawata} D.,  {Miglio} A.,  {Davies} G.~R.,   {Grand} R.
  J.~J.,  2021, \mn@doi [\mnras] {10.1093/mnras/stab639}, \href
  {https://ui.adsabs.harvard.edu/abs/2021MNRAS.503.2814C} {503, 2814}

\bibitem[\protect\citeauthoryear{{Clarke} \& {Gerhard}}{{Clarke} \&
  {Gerhard}}{2021}]{Clarke+Gerhard21}
{Clarke} J.,  {Gerhard} O.,  2021, arXiv e-prints, \href
  {https://ui.adsabs.harvard.edu/abs/2021arXiv210710875C} {p. arXiv:2107.10875}

\bibitem[\protect\citeauthoryear{{Cropper} et~al.,}{{Cropper}
  et~al.}{2018}]{Cropper+18}
{Cropper} M.,  et~al., 2018, \mn@doi [\aap] {10.1051/0004-6361/201832763},
  \href {https://ui.adsabs.harvard.edu/abs/2018A&A...616A...5C} {616, A5}

\bibitem[\protect\citeauthoryear{{D'Onghia} \& {L. Aguerri}}{{D'Onghia} \& {L.
  Aguerri}}{2020}]{D'Onghia+Aguerri20}
{D'Onghia} E.,  {L. Aguerri} J.~A.,  2020, \mn@doi [\apj]
  {10.3847/1538-4357/ab6bd6}, \href
  {https://ui.adsabs.harvard.edu/abs/2020ApJ...890..117D} {890, 117}

\bibitem[\protect\citeauthoryear{{De Simone}, {Wu}  \& {Tremaine}}{{De Simone}
  et~al.}{2004}]{DeSimone+Wu+Tremaine04}
{De Simone} R.,  {Wu} X.,   {Tremaine} S.,  2004, \mn@doi [\mnras]
  {10.1111/j.1365-2966.2004.07675.x}, \href
  {http://adsabs.harvard.edu/abs/2004MNRAS.350..627D} {350, 627}

\bibitem[\protect\citeauthoryear{{Dehnen}}{{Dehnen}}{1999}]{Dehnen99}
{Dehnen} W.,  1999, \mn@doi [\apjl] {10.1086/312299}, \href
  {https://ui.adsabs.harvard.edu/abs/1999ApJ...524L..35D} {524, L35}

\bibitem[\protect\citeauthoryear{{Dehnen}}{{Dehnen}}{2000}]{wd00}
{Dehnen} W.,  2000, \mn@doi [\aj] {10.1086/301226}, \href
  {http://adsabs.harvard.edu/abs/2000AJ....119..800D} {119, 800}

\bibitem[\protect\citeauthoryear{{Fragkoudi} et~al.,}{{Fragkoudi}
  et~al.}{2019}]{Fragkoudi+19}
{Fragkoudi} F.,  et~al., 2019, \mn@doi [\mnras] {10.1093/mnras/stz1875}, \href
  {https://ui.adsabs.harvard.edu/abs/2019MNRAS.488.3324F} {488, 3324}

\bibitem[\protect\citeauthoryear{{Fragkoudi} et~al.,}{{Fragkoudi}
  et~al.}{2020}]{Fragkoudi+Grand+Pakmor+20}
{Fragkoudi} F.,  et~al., 2020, \mn@doi [\mnras] {10.1093/mnras/staa1104}, \href
  {https://ui.adsabs.harvard.edu/abs/2020MNRAS.494.5936F} {494, 5936}

\bibitem[\protect\citeauthoryear{{Freeman} et~al.,}{{Freeman}
  et~al.}{2013}]{Freeman+Ness+13}
{Freeman} K.,  et~al., 2013, \mn@doi [\mnras] {10.1093/mnras/sts305}, \href
  {https://ui.adsabs.harvard.edu/abs/2013MNRAS.428.3660F} {428, 3660}

\bibitem[\protect\citeauthoryear{{Friske} \& {Sch{\"o}nrich}}{{Friske} \&
  {Sch{\"o}nrich}}{2019}]{Friske+Schoenrich19}
{Friske} J. K.~S.,  {Sch{\"o}nrich} R.,  2019, \mn@doi [\mnras]
  {10.1093/mnras/stz2951}, \href
  {https://ui.adsabs.harvard.edu/abs/2019MNRAS.490.5414F} {490, 5414}

\bibitem[\protect\citeauthoryear{{Fujii}, {B{\'e}dorf}, {Baba}  \& {Portegies
  Zwart}}{{Fujii} et~al.}{2019}]{Fujii+2019}
{Fujii} M.~S.,  {B{\'e}dorf} J.,  {Baba} J.,   {Portegies Zwart} S.,  2019,
  \mn@doi [\mnras] {10.1093/mnras/sty2747}, \href
  {http://adsabs.harvard.edu/abs/2019MNRAS.482.1983F} {482, 1983}

\bibitem[\protect\citeauthoryear{{Gaia Collaboration} et~al.,}{{Gaia
  Collaboration} et~al.}{2018a}]{Gaia+Brown+18}
{Gaia Collaboration} et~al., 2018a, \mn@doi [\aap]
  {10.1051/0004-6361/201833051}, \href
  {https://ui.adsabs.harvard.edu/abs/2018A&A...616A...1G} {616, A1}

\bibitem[\protect\citeauthoryear{{Gaia Collaboration} et~al.,}{{Gaia
  Collaboration} et~al.}{2018b}]{Gaia+Katz18Disc}
{Gaia Collaboration} et~al., 2018b, \mn@doi [\aap]
  {10.1051/0004-6361/201832865}, \href
  {https://ui.adsabs.harvard.edu/abs/2018A&A...616A..11G} {616, A11}

\bibitem[\protect\citeauthoryear{{Gaia Collaboration} et~al.,}{{Gaia
  Collaboration} et~al.}{2021a}]{Gaia+Brown+21}
{Gaia Collaboration} et~al., 2021a, \mn@doi [\aap]
  {10.1051/0004-6361/202039657}, \href
  {https://ui.adsabs.harvard.edu/abs/2021A&A...649A...1G} {649, A1}

\bibitem[\protect\citeauthoryear{{Gaia Collaboration} et~al.,}{{Gaia
  Collaboration} et~al.}{2021b}]{Gaia+Antoja+21}
{Gaia Collaboration} et~al., 2021b, \mn@doi [\aap]
  {10.1051/0004-6361/202039714}, \href
  {https://ui.adsabs.harvard.edu/abs/2021A&A...649A...8G} {649, A8}

\bibitem[\protect\citeauthoryear{{Gouda} \& {Jasmine Team}}{{Gouda} \& {Jasmine
  Team}}{2020}]{Gouda+20}
{Gouda} N.,  {Jasmine Team} 2020, in {Valluri} M.,  {Sellwood} J.~A.,  eds,
  IAU Symposium Vol. 353, Galactic Dynamics in the Era of Large Surveys. pp
  51--53, \mn@doi{10.1017/S1743921319007968}

\bibitem[\protect\citeauthoryear{{Grand}, {Kawata}  \& {Cropper}}{{Grand}
  et~al.}{2012}]{gkc12a}
{Grand} R.~J.~J.,  {Kawata} D.,   {Cropper} M.,  2012, \mn@doi [\mnras]
  {10.1111/j.1365-2966.2012.20411.x}, \href
  {http://adsabs.harvard.edu/abs/2012MNRAS.421.1529G} {421, 1529}

\bibitem[\protect\citeauthoryear{{Grand}, {Kawata}  \& {Cropper}}{{Grand}
  et~al.}{2015}]{gkc15}
{Grand} R.~J.~J.,  {Kawata} D.,   {Cropper} M.,  2015, \mn@doi [\mnras]
  {10.1093/mnras/stv016}, \href
  {http://adsabs.harvard.edu/abs/2015MNRAS.447.4018G} {447, 4018}

\bibitem[\protect\citeauthoryear{{Grand} et~al.,}{{Grand}
  et~al.}{2016}]{Grand+Springel+Kawata+16}
{Grand} R. J.~J.,  et~al., 2016, \mn@doi [\mnras] {10.1093/mnrasl/slw086},
  \href {https://ui.adsabs.harvard.edu/abs/2016MNRAS.460L..94G} {460, L94}

\bibitem[\protect\citeauthoryear{{Gravity Collaboration} et~al.,}{{Gravity
  Collaboration} et~al.}{2019}]{Gravity+GCdistance19}
{Gravity Collaboration} et~al., 2019, \mn@doi [\aap]
  {10.1051/0004-6361/201935656}, \href
  {https://ui.adsabs.harvard.edu/abs/2019A&A...625L..10G} {625, L10}

\bibitem[\protect\citeauthoryear{{Hilmi} et~al.,}{{Hilmi}
  et~al.}{2020}]{Hilmi+Minchev+Buck+20}
{Hilmi} T.,  et~al., 2020, \mn@doi [\mnras] {10.1093/mnras/staa1934}, \href
  {https://ui.adsabs.harvard.edu/abs/2020MNRAS.497..933H} {497, 933}

\bibitem[\protect\citeauthoryear{{Hunt} \& {Bovy}}{{Hunt} \&
  {Bovy}}{2018}]{Hunt+Bovy18}
{Hunt} J.~A.~S.,  {Bovy} J.,  2018, \mn@doi [\mnras] {10.1093/mnras/sty921},
  \href {http://adsabs.harvard.edu/abs/2018MNRAS.477.3945H} {477, 3945}

\bibitem[\protect\citeauthoryear{{Hunt}, {Kawata}, {Monari}, {Grand}, {Famaey}
  \& {Siebert}}{{Hunt} et~al.}{2017}]{Hunt+17}
{Hunt} J.~A.~S.,  {Kawata} D.,  {Monari} G.,  {Grand} R.~J.~J.,  {Famaey} B.,
  {Siebert} A.,  2017, \mn@doi [\mnras] {10.1093/mnrasl/slw257}, \href
  {http://adsabs.harvard.edu/abs/2017MNRAS.467L..21H} {467, L21}

\bibitem[\protect\citeauthoryear{{Hunt}, {Hong}, {Bovy}, {Kawata}  \&
  {Grand}}{{Hunt} et~al.}{2018}]{Hunt+Hong+Bovy+18}
{Hunt} J.~A.~S.,  {Hong} J.,  {Bovy} J.,  {Kawata} D.,   {Grand} R.~J.~J.,
  2018, \mn@doi [\mnras] {10.1093/mnras/sty2532}, \href
  {http://adsabs.harvard.edu/abs/2018MNRAS.481.3794H} {481, 3794}

\bibitem[\protect\citeauthoryear{{Hunt}, {Bub}, {Bovy}, {Mackereth}, {Trick}
  \& {Kawata}}{{Hunt} et~al.}{2019}]{Hunt+Bub+Bovy+19}
{Hunt} J. A.~S.,  {Bub} M.~W.,  {Bovy} J.,  {Mackereth} J.~T.,  {Trick} W.~H.,
   {Kawata} D.,  2019, \mn@doi [\mnras] {10.1093/mnras/stz2667}, \href
  {https://ui.adsabs.harvard.edu/abs/2019MNRAS.490.1026H} {490, 1026}

\bibitem[\protect\citeauthoryear{{Hunt}, {Johnston}, {Pettitt}, {Cunningham},
  {Kawata}  \& {Hogg}}{{Hunt} et~al.}{2020}]{Hunt+Johnston+Pettitt+20}
{Hunt} J. A.~S.,  {Johnston} K.~V.,  {Pettitt} A.~R.,  {Cunningham} E.~C.,
  {Kawata} D.,   {Hogg} D.~W.,  2020, \mn@doi [\mnras]
  {10.1093/mnras/staa1987}, \href
  {https://ui.adsabs.harvard.edu/abs/2020MNRAS.497..818H} {497, 818}

\bibitem[\protect\citeauthoryear{{Irrgang}, {Wilcox}, {Tucker}  \&
  {Schiefelbein}}{{Irrgang} et~al.}{2013}]{Irrgang+13}
{Irrgang} A.,  {Wilcox} B.,  {Tucker} E.,   {Schiefelbein} L.,  2013, \mn@doi
  [\aap] {10.1051/0004-6361/201220540}, \href
  {https://ui.adsabs.harvard.edu/abs/2013A&A...549A.137I} {549, A137}

\bibitem[\protect\citeauthoryear{{Kawata}, {Hunt}, {Grand}, {Pasetto}  \&
  {Cropper}}{{Kawata} et~al.}{2014}]{khgpc14}
{Kawata} D.,  {Hunt} J.~A.~S.,  {Grand} R.~J.~J.,  {Pasetto} S.,   {Cropper}
  M.,  2014, \mn@doi [\mnras] {10.1093/mnras/stu1292}, \href
  {http://adsabs.harvard.edu/abs/2014MNRAS.443.2757K} {443, 2757}

\bibitem[\protect\citeauthoryear{{Kawata} et~al.,}{{Kawata}
  et~al.}{2018}]{Kawata+18}
{Kawata} D.,  et~al., 2018, \mn@doi [\mnras] {10.1093/mnras/stx2464}, \href
  {http://adsabs.harvard.edu/abs/2018MNRAS.473..867K} {473, 867}

\bibitem[\protect\citeauthoryear{{Khanna} et~al.,}{{Khanna}
  et~al.}{2019}]{Khanna+19}
{Khanna} S.,  et~al., 2019, \mn@doi [\mnras] {10.1093/mnras/stz2462}, \href
  {https://ui.adsabs.harvard.edu/abs/2019MNRAS.489.4962K} {489, 4962}

\bibitem[\protect\citeauthoryear{{Khoperskov}, {Di Matteo}, {Gerhard}, {Katz},
  {Haywood}, {Combes}, {Berczik}  \& {Gomez}}{{Khoperskov}
  et~al.}{2019}]{Khoperskov+DiMatteo+Gerhard+19}
{Khoperskov} S.,  {Di Matteo} P.,  {Gerhard} O.,  {Katz} D.,  {Haywood} M.,
  {Combes} F.,  {Berczik} P.,   {Gomez} A.,  2019, \mn@doi [\aap]
  {10.1051/0004-6361/201834707}, \href
  {https://ui.adsabs.harvard.edu/abs/2019A&A...622L...6K} {622, L6}

\bibitem[\protect\citeauthoryear{{Kunder} et~al.,}{{Kunder}
  et~al.}{2012}]{Kunder+Koch+Rich+12}
{Kunder} A.,  et~al., 2012, \mn@doi [\aj] {10.1088/0004-6256/143/3/57}, \href
  {https://ui.adsabs.harvard.edu/abs/2012AJ....143...57K} {143, 57}

\bibitem[\protect\citeauthoryear{{Laporte}, {Minchev}, {Johnston}  \&
  {G{\'o}mez}}{{Laporte} et~al.}{2019}]{LMJG19}
{Laporte} C. F.~P.,  {Minchev} I.,  {Johnston} K.~V.,   {G{\'o}mez} F.~A.,
  2019, \mn@doi [\mnras] {10.1093/mnras/stz583}, \href
  {https://ui.adsabs.harvard.edu/abs/2019MNRAS.485.3134L} {485, 3134}

\bibitem[\protect\citeauthoryear{{Lindegren} et~al.,}{{Lindegren}
  et~al.}{2021}]{Lindegren+Bastian+Biermann+21}
{Lindegren} L.,  et~al., 2021, \mn@doi [\aap] {10.1051/0004-6361/202039653},
  \href {https://ui.adsabs.harvard.edu/abs/2021A&A...649A...4L} {649, A4}

\bibitem[\protect\citeauthoryear{{Lynden-Bell} \& {Kalnajs}}{{Lynden-Bell} \&
  {Kalnajs}}{1972}]{lbk72}
{Lynden-Bell} D.,  {Kalnajs} A.~J.,  1972, \mn@doi [\mnras]
  {10.1093/mnras/157.1.1}, \href
  {http://adsabs.harvard.edu/abs/1972MNRAS.157....1L} {157, 1}

\bibitem[\protect\citeauthoryear{{Majewski} et~al.,}{{Majewski}
  et~al.}{2017}]{Majewski+Schiavon+Frinchaboy+17}
{Majewski} S.~R.,  et~al., 2017, \mn@doi [\aj] {10.3847/1538-3881/aa784d},
  \href {https://ui.adsabs.harvard.edu/abs/2017AJ....154...94M} {154, 94}

\bibitem[\protect\citeauthoryear{{McMillan}}{{McMillan}}{2017}]{McMillan17}
{McMillan} P.~J.,  2017, \mn@doi [\mnras] {10.1093/mnras/stw2759}, \href
  {https://ui.adsabs.harvard.edu/abs/2017MNRAS.465...76M} {465, 76}

\bibitem[\protect\citeauthoryear{{Minniti} et~al.,}{{Minniti}
  et~al.}{2010}]{Minniti+Lucas+Emerson+10}
{Minniti} D.,  et~al., 2010, \mn@doi [\na] {10.1016/j.newast.2009.12.002},
  \href {https://ui.adsabs.harvard.edu/abs/2010NewA...15..433M} {15, 433}

\bibitem[\protect\citeauthoryear{{Monari}, {Kawata}, {Hunt}  \&
  {Famaey}}{{Monari} et~al.}{2017a}]{Monari+17}
{Monari} G.,  {Kawata} D.,  {Hunt} J.~A.~S.,   {Famaey} B.,  2017a, \mn@doi
  [\mnras] {10.1093/mnrasl/slw238}, \href
  {http://adsabs.harvard.edu/abs/2017MNRAS.466L.113M} {466, L113}

\bibitem[\protect\citeauthoryear{{Monari}, {Famaey}, {Fouvry}  \&
  {Binney}}{{Monari} et~al.}{2017b}]{Monari+Famaey+Fouvry+Binney17}
{Monari} G.,  {Famaey} B.,  {Fouvry} J.-B.,   {Binney} J.,  2017b, \mn@doi
  [\mnras] {10.1093/mnras/stx1825}, \href
  {https://ui.adsabs.harvard.edu/abs/2017MNRAS.471.4314M} {471, 4314}

\bibitem[\protect\citeauthoryear{{Monari}, {Famaey}, {Siebert}, {Wegg}  \&
  {Gerhard}}{{Monari} et~al.}{2019}]{Monari+Famaey+Siebert+Wegg+Gerhard19}
{Monari} G.,  {Famaey} B.,  {Siebert} A.,  {Wegg} C.,   {Gerhard} O.,  2019,
  \mn@doi [\aap] {10.1051/0004-6361/201834820}, \href
  {https://ui.adsabs.harvard.edu/abs/2019A&A...626A..41M} {626, A41}

\bibitem[\protect\citeauthoryear{{Ness} et~al.,}{{Ness}
  et~al.}{2013}]{Ness+Freeman+Athanassoula+13}
{Ness} M.,  et~al., 2013, \mn@doi [\mnras] {10.1093/mnras/sts629}, \href
  {https://ui.adsabs.harvard.edu/abs/2013MNRAS.430..836N} {430, 836}

\bibitem[\protect\citeauthoryear{Pedregosa et~al.,}{Pedregosa
  et~al.}{2011}]{scikit-learn}
Pedregosa F.,  et~al., 2011, Journal of Machine Learning Research, 12, 2825

\bibitem[\protect\citeauthoryear{{P{\'e}rez-Villegas}, {Portail}, {Wegg}  \&
  {Gerhard}}{{P{\'e}rez-Villegas} et~al.}{2017}]{Perez-Villegas+17}
{P{\'e}rez-Villegas} A.,  {Portail} M.,  {Wegg} C.,   {Gerhard} O.,  2017,
  \mn@doi [\apjl] {10.3847/2041-8213/aa6c26}, \href
  {https://ui.adsabs.harvard.edu/abs/2017ApJ...840L...2P} {840, L2}

\bibitem[\protect\citeauthoryear{{Portail}, {Wegg}, {Gerhard}  \&
  {Martinez-Valpuesta}}{{Portail} et~al.}{2015}]{Portail+15}
{Portail} M.,  {Wegg} C.,  {Gerhard} O.,   {Martinez-Valpuesta} I.,  2015,
  \mn@doi [\mnras] {10.1093/mnras/stv058}, \href
  {http://adsabs.harvard.edu/abs/2015MNRAS.448..713P} {448, 713}

\bibitem[\protect\citeauthoryear{{Quillen}, {Dougherty}, {Bagley}, {Minchev}
  \& {Comparetta}}{{Quillen} et~al.}{2011}]{qdbmc11}
{Quillen} A.~C.,  {Dougherty} J.,  {Bagley} M.~B.,  {Minchev} I.,
  {Comparetta} J.,  2011, \mn@doi [\mnras] {10.1111/j.1365-2966.2011.19349.x},
  \href {http://adsabs.harvard.edu/abs/2011MNRAS.417..762Q} {417, 762}

\bibitem[\protect\citeauthoryear{{Ramos}, {Antoja}  \& {Figueras}}{{Ramos}
  et~al.}{2018}]{Ramos+18}
{Ramos} P.,  {Antoja} T.,   {Figueras} F.,  2018, \mn@doi [\aap]
  {10.1051/0004-6361/201833494}, \href
  {https://ui.adsabs.harvard.edu/abs/2018A&A...619A..72R} {619, A72}

\bibitem[\protect\citeauthoryear{{Reid} \& {Brunthaler}}{{Reid} \&
  {Brunthaler}}{2020}]{Reid+Brunthaler20}
{Reid} M.~J.,  {Brunthaler} A.,  2020, \mn@doi [\apj]
  {10.3847/1538-4357/ab76cd}, \href
  {https://ui.adsabs.harvard.edu/abs/2020ApJ...892...39R} {892, 39}

\bibitem[\protect\citeauthoryear{{Saitoh}, {Daisaka}, {Kokubo}, {Makino},
  {Okamoto}, {Tomisaka}, {Wada}  \& {Yoshida}}{{Saitoh}
  et~al.}{2008}]{Saitoh+2008}
{Saitoh} T.~R.,  {Daisaka} H.,  {Kokubo} E.,  {Makino} J.,  {Okamoto} T.,
  {Tomisaka} K.,  {Wada} K.,   {Yoshida} N.,  2008, \pasj, \href
  {http://adsabs.harvard.edu/abs/2008PASJ...60..667S} {60, 667}

\bibitem[\protect\citeauthoryear{{Sanders}, {Smith}  \& {Evans}}{{Sanders}
  et~al.}{2019}]{Sanders+Smith+Evans19}
{Sanders} J.~L.,  {Smith} L.,   {Evans} N.~W.,  2019, \mn@doi [\mnras]
  {10.1093/mnras/stz1827}, \href
  {https://ui.adsabs.harvard.edu/abs/2019MNRAS.488.4552S} {488, 4552}

\bibitem[\protect\citeauthoryear{{Sch{\"o}nrich}, {McMillan}  \&
  {Eyer}}{{Sch{\"o}nrich} et~al.}{2019}]{Schoenrich+McMillan+Eyer19}
{Sch{\"o}nrich} R.,  {McMillan} P.,   {Eyer} L.,  2019, \mn@doi [\mnras]
  {10.1093/mnras/stz1451}, \href
  {https://ui.adsabs.harvard.edu/abs/2019MNRAS.487.3568S} {487, 3568}

\bibitem[\protect\citeauthoryear{{Sellwood}}{{Sellwood}}{2010}]{jas10}
{Sellwood} J.~A.,  2010, \mn@doi [\mnras] {10.1111/j.1365-2966.2010.17305.x},
  \href {http://adsabs.harvard.edu/abs/2010MNRAS.409..145S} {409, 145}

\bibitem[\protect\citeauthoryear{{Sellwood} \& {Binney}}{{Sellwood} \&
  {Binney}}{2002}]{jsjb02}
{Sellwood} J.~A.,  {Binney} J.~J.,  2002, \mn@doi [\mnras]
  {10.1046/j.1365-8711.2002.05806.x}, \href
  {http://adsabs.harvard.edu/abs/2002MNRAS.336..785S} {336, 785}

\bibitem[\protect\citeauthoryear{{Shen}, {Rich}, {Kormendy}, {Howard}, {De
  Propris}  \& {Kunder}}{{Shen} et~al.}{2010}]{Shen+Rich+Kormendy+10}
{Shen} J.,  {Rich} R.~M.,  {Kormendy} J.,  {Howard} C.~D.,  {De Propris} R.,
  {Kunder} A.,  2010, \mn@doi [\apjl] {10.1088/2041-8205/720/1/L72}, \href
  {https://ui.adsabs.harvard.edu/abs/2010ApJ...720L..72S} {720, L72}

\bibitem[\protect\citeauthoryear{{Solway}, {Sellwood}  \&
  {Sch{\"o}nrich}}{{Solway} et~al.}{2012}]{sss12}
{Solway} M.,  {Sellwood} J.~A.,   {Sch{\"o}nrich} R.,  2012, \mn@doi [\mnras]
  {10.1111/j.1365-2966.2012.20712.x}, \href
  {http://ukads.nottingham.ac.uk/abs/2012MNRAS.422.1363S} {422, 1363}

\bibitem[\protect\citeauthoryear{{Sormani}, {Binney}  \& {Magorrian}}{{Sormani}
  et~al.}{2015}]{Sormani+15c}
{Sormani} M.~C.,  {Binney} J.,   {Magorrian} J.,  2015, \mn@doi [\mnras]
  {10.1093/mnras/stv2067}, \href
  {https://ui.adsabs.harvard.edu/abs/2015MNRAS.454.1818S} {454, 1818}

\bibitem[\protect\citeauthoryear{{Stephens} \& {Boesgaard}}{{Stephens} \&
  {Boesgaard}}{2002}]{sb02}
{Stephens} A.,  {Boesgaard} A.~M.,  2002, \mn@doi [\aj] {10.1086/338898}, \href
  {http://adsabs.harvard.edu/cgi-bin/nph-bib_query?bibcode=2002AJ....123.1647S&db_key=AST}
  {123, 1647}

\bibitem[\protect\citeauthoryear{{Trick}}{{Trick}}{2020}]{Trick21}
{Trick} W.~H.,  2020, arXiv e-prints, \href
  {https://ui.adsabs.harvard.edu/abs/2020arXiv201101233T} {p. arXiv:2011.01233}

\bibitem[\protect\citeauthoryear{{Trick}, {Coronado}  \& {Rix}}{{Trick}
  et~al.}{2019}]{Trick+Coronado+Rix+19}
{Trick} W.~H.,  {Coronado} J.,   {Rix} H.-W.,  2019, \mn@doi [\mnras]
  {10.1093/mnras/stz209}, \href
  {https://ui.adsabs.harvard.edu/abs/2019MNRAS.484.3291T} {484, 3291}

\bibitem[\protect\citeauthoryear{{Trick}, {Fragkoudi}, {Hunt}, {Mackereth}  \&
  {White}}{{Trick} et~al.}{2021}]{Trick+Fragkoudi+Hunt+21}
{Trick} W.~H.,  {Fragkoudi} F.,  {Hunt} J. A.~S.,  {Mackereth} J.~T.,   {White}
  S. D.~M.,  2021, \mn@doi [\mnras] {10.1093/mnras/staa3317}, \href
  {https://ui.adsabs.harvard.edu/abs/2021MNRAS.500.2645T} {500, 2645}

\bibitem[\protect\citeauthoryear{{Vasiliev}}{{Vasiliev}}{2019}]{Vasiliev_AGAMA19}
{Vasiliev} E.,  2019, \mn@doi [\mnras] {10.1093/mnras/sty2672}, \href
  {https://ui.adsabs.harvard.edu/abs/2019MNRAS.482.1525V} {482, 1525}

\bibitem[\protect\citeauthoryear{{Wu}, {Pfenniger}  \& {Taam}}{{Wu}
  et~al.}{2016}]{Wu+Pfenniger+Taam16}
{Wu} Y.-T.,  {Pfenniger} D.,   {Taam} R.~E.,  2016, \mn@doi [\apj]
  {10.3847/0004-637X/830/2/111}, \href
  {https://ui.adsabs.harvard.edu/abs/2016ApJ...830..111W} {830, 111}

\makeatother
\end{thebibliography}

%\begin{thebibliography}{99}
%\end{thebibliography}

%%%%%%%%%%%%%%%%%%%%%%%%%%%%%%%%%%%%%%%%%%%%%%%%%%

%%%%%%%%%%%%%%%%% APPENDICES %%%%%%%%%%%%%%%%%%%%%

\appendix

%%%%%%%%%%%%%%%%%%%%%%%%%%%%%%%%%%%%%%%%%%%%%%%%%%

% Don't change these lines
\bsp	% typesetting comment
\label{lastpage}
\end{document}